\documentclass[journal]{IEEEtran}
\usepackage{cite}
\usepackage[pdftex]{graphicx}
\usepackage{hyperref}
\usepackage{footnote} 
\usepackage{amsmath, amsfonts, amssymb, amsthm}
\allowdisplaybreaks
\theoremstyle{plain}

\newtheorem{assumption}{Assumption}
\newtheorem{definition}{Definition}
\newtheorem{lemma}{Lemma}

\newtheorem{corollary}{Corollary}
\newtheorem{theorem}{Theorem}

\usepackage{tikz}\usetikzlibrary{calc, positioning}
\usepackage[linesnumbered, lined, boxed]{algorithm2e}
\usepackage{lineno}\modulolinenumbers[5]

\begin{document}
\title{Fast and Stable Nonconvex Constrained Distributed Optimization: The ELLADA Algorithm}
\author{Wentao~Tang~and~Prodromos~Daoutidis
\thanks{The authors are with the Department of Chemical Engineering and Materials Science, University of Minnesota, Minneapolis, MN 55455, U.S.A. (e-mails: tangx647@umn.edu, daout001@umn.edu).}}
\maketitle
\begin{abstract}
	Distributed optimization, where the computations are performed in a localized and coordinated manner using multiple agents, is a promising approach for solving large-scale optimization problems, e.g., those arising in model predictive control (MPC) of large-scale plants. However, a distributed optimization algorithm that is computationally efficient, globally convergent, amenable to nonconvex constraints and general inter-subsystem interactions remains an open problem. In this paper, we combine three important modifications to the classical alternating direction method of multipliers (ADMM) for distributed optimization. Specifically, (i) an extra-layer architecture is adopted to accommodate nonconvexity and handle inequality constraints, (ii) equality-constrained nonlinear programming (NLP) problems are allowed to be solved approximately, and (iii) a modified Anderson acceleration is employed for reducing the number of iterations. Theoretical convergence towards stationary solutions and computational complexity of the proposed algorithm, named ELLADA, is established. Its application to distributed nonlinear MPC is also described and illustrated through a benchmark process system. 
\end{abstract}
\IEEEpeerreviewmaketitle

\section{Introduction}
\IEEEPARstart{D}{istributed} optimization \cite{boyd2011distributed, nedic2014distributed, molzahn2017survey} refers to methods of performing optimization using a distributed architecture -- multiple networked agents are used for subsystems and necessary information among the agents is communicated to coordinate the distributed computation. An important desirable application of distributed optimization is in model predictive control (MPC), where control decisions are made through solving an optimal control problem minimizing the cost associated with the predicted trajectory in a future horizon subject to the system dynamics and operational constraints \cite{rawlings2018MPC}. For large-scale systems, it is desirable to seek a decomposition (e.g., using community detection or network block structures \cite{daoutidis2018decomposing, tang2018optimal, daoutidis2019decomposition}) and deploy distributed MPC strategies \cite{scattolini2009architectures, christofides2013distributed, negenborn2014distributed}, which allows better performance than fully decentralized MPC by enabling coordination, while avoiding assembling and computing on a monolithic model in centralized MPC. 

\par Despite efficient algorithms for solving monolithic nonlinear programming (NLP) problems in centralized MPC (e.g., \cite{patterson2014gpops, mao2016successive, biegler2018large}), extending them into distributed algorithms is nontrivial. A typical approach of distributed MPC is to iterate the control inputs among the subsystems (in sequence or in parallel) \cite{stewart2010cooperative, liu2010sequential, chen2012distributed}. The input iteration routine is typically either semi-decentralized by implicitly assuming that the subsystems interact only through inputs and considering state coupling as disturbances, or semi-centralized by using moving-horizon predictions based on the entire system, which, however, contradicts the fact that the subsystem models should be usually packaged inside the local agents rather than shared over the entire system. Distributed MPC under truly localized model information is typically restricted to linear systems \cite{venkat2005stability, farina2012distributed, giselsson2013accelerated}.

\par
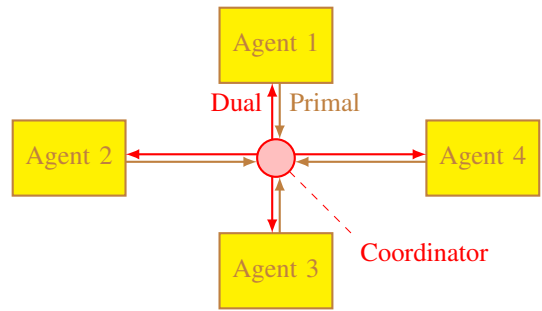
\begin{figure}
	\centering
	\begin{tikzpicture}
	\draw[thick, red, fill=pink] (0, 0) circle [radius = 0.25]; 
	\draw[-, dashed, red] (0.1768, -0.1768) -- (1, -1) node[below right, red]{Coordinator}; 
	
	\draw[thick, brown, fill=yellow] (-0.75, 1) rectangle (0.75, 2);
	\node[brown] at (0, 1.5) {Agent 1}; 
	\draw[thick, red, -latex] (-0.05, 0.25) -- (-0.05, 1) node[red, below left]{Dual};
	\draw[thick, brown, -latex] (0.05, 1) node[brown, below right]{Primal} -- (0.05, 0.25);
	
	\draw[thick, brown, fill=yellow] (-3.5, -0.5) rectangle (-2, 0.5);
	\node[brown] at (-2.75, 0) {Agent 2}; 
	\draw[thick, red, -latex] (-0.25, 0.05) -- (-2, 0.05);
	\draw[thick, brown, -latex] (-2, -0.05) -- (-0.25, -0.05);
	
	\draw[thick, brown, fill=yellow] (2, -0.5) rectangle (3.5, 0.5);
	\node[brown] at (2.75, 0) {Agent 4};
	\draw[thick, red, -latex] (0.25, 0.05) -- (2, 0.05);
	\draw[thick, brown, -latex] (2, -0.05) -- (0.25, -0.05);
	
	\draw[thick, brown, fill=yellow] (-0.75, -1) rectangle (0.75, -2);
	\node[brown] at (0, -1.5) {Agent 3}; 
	\draw[thick, red, -latex] (-0.05, -0.25) -- (-0.05, -1);
	\draw[thick, brown, -latex] (0.05, -1) -- (0.05, -0.25);
	\end{tikzpicture}
	\caption{Primal-dual distributed optimization.}\label{fig:primal-dual}
\end{figure}

 We note that in general, distributed nonlinear MPC with subsystem interactions should be considered as a \emph{distributed optimization problem under nonconvex constraints}. To solve such problems using distributed agents with local model knowledge, Lagrangian decomposition using dual relaxation of complicating interactions \cite[Section~9]{grune2017distributed} and the alternating direction method of multipliers (ADMM) algorithm using augmented Lagrangian \cite{farokhi2014distributed, mota2014distributed} were proposed as general frameworks. These are primal-dual iterative algorithms. As illustrated in Fig. \ref{fig:primal-dual}, in each iteration, the distributed agents receive the dual information from the coordinator and execute subroutines to solve their own subproblems, and a coordinator collects information of their solutions to update the duals. 

\par \emph{Convergence} is the most basic requirement but also a major challenge in distributed optimization under nonconvex constraints. Although distributed optimization with nonconvex objective functions has been well discussed \cite{magnusson2015convergence, tatarenko2017non, chatzipanagiotis2017convergence, wang2019global}, the nonconvex constraints appear much more difficult to handle.
To guarantee convergence, \cite{hours2015parametric} suggested dualizing and penalizing all nonconvex constraints, making them undifferentiated and tractable by ADMM; however, this alteration of the problem structure eliminates the option for distributed agents to use any subroutine other than the method of multipliers (MM). \cite{houska2016augmented} used a quadratic programming problem to decide the dual variables in the augmented Lagrangian as well as an extrapolation of primal updates; this algorithm, however, involves a central agent that extracts Hessian and gradient information of the subsystem models from the distributed agents in every iteration, and is thus essentially semi-centralized. \cite{scutari2016parallel} adopted feasibility-preserving convex approximations to approach the solution, which is applicable to problems without nonconvex equality constraints. We note that several recent papers (e.g., \cite{sun2019two, jiang2019structured, yang2020proximal}) proposed the idea of placing slack variables corresponding to the inter-subsystem constraints and forcing the decay to zero by tightening the penalty parameters of slack variables. This modification to the ADMM with slack variables and their penalties leads to a globally convergent \emph{extra-layer} augmented Lagrangian-based algorithm with preserved agent-coordinator problem architecture. 

\par \emph{Computational efficiency} is also of critical importance for distributed optimization, especially in MPC. The slothfulness of primal-dual algorithms typically arises from two issues. First, the subgradient (first-order) update of dual variables restricts the number of iterations to be of linear complexity \cite{lin2015global, hong2017linear, makhdoumi2017convergence}. For convex problems, momentum methods \cite{goldstein2014fast, ouyang2015accelerated} can be adopted to obtain second-order dual updates. Such momentum acceleration can not be directly extended to nonconvex problems without a positive definite curvature, although our previous numerical study showed that a discounted momentum may allow limited improvement \cite{tang2019accelerated}. Another effort to accelerate dual updates in convex ADMM is based on Krylov subspace methods \cite{zhang2018gmres}. Under nonconvexity, it was only very recently realized that \emph{Anderson acceleration}, a multi-secant technique for fixed-point problems, can be generally used to accelerate the dual variables \cite{zhang2018globally, fu2019anderson, zhang2019accelerating}. 

\par The second cause for the high computational cost of distributed optimization is the instruction on the distributed agents to fully solve their subproblems to high precision in each iteration. Such exhaustive efforts may be unnecessary since the dual information to be received from the coordinator will keep changing. For convex problems, it is possible to linearize the augmented Lagrangian and replace the distributed subproblems with Arrow-Hurwicz-Uzawa gradient flows \cite{dhingra2018proximal}. In the presence of nonconvexity of the objective functions, a dual perturbation technique to restore the convergence of the augmented Lagrangian was proposed in \cite{hajinezhad2019perturbed}. It is yet unknown how to accommodate such gradient flows to nonconvex constraints. A different approach is to allow inexact solution of the subproblems with adaptively tightening tolerances \cite{eckstein2017approximate}. Such an \emph{approximate ADMM algorithm} allows a better balance between the primal and dual updates, and avoids wasteful computational steps inside the subroutines. 

\par The purpose of this work is to develop a convergent and computationally efficient algorithm for distributed optimization under nonconvex constraints. Although the algorithm is in principle not restricted to any specific problem, we consider the implementation of distributed nonlinear MPC as an important application. Based on the above discussion, we identify the following modifications to the classical ADMM algorithm as the key to mitigating the challenges in convergence and computational complexity: (i) additional slack variables are placed on the constraints relating the distributed agents and the coordinator, (ii) approximate optimization is performed in the distributed agents, and (iii) the Anderson acceleration technique is adopted by the coordinator. We therefore combine and extend as appropriate these techniques into a new algorithm with a two-layer augmented Lagrangian-based architecture, in which the outer layer handles the slack variables as well as inequality constraints by using a barrier technique, and the inner layer performs approximate ADMM under an acceleration scheme. With guaranteed stability and elevated speed, \emph{to the best knowledge of the authors, the proposed algorithm is the first practical and generic algorithm of its kind for distributed nonlinear MPC with truly localized model information.} We name this algorithm as \textbf{ELLADA} (standing for \textbf{e}xtra-\textbf{l}ayer augmented \textbf{L}agrangian-based \textbf{a}ccelerated \textbf{d}istributed \textbf{a}pproximate optimization). 

\par The paper discusses the movitation, develops the ELLADA algorithm and establishes its theoretical properties. An application to a benchmark quadruple tank process is also presented. The remainder of this paper is organized as follows. In Section \ref{sec:ADMM}, we first review the classical ADMM and its modified versions. Then we derive our ELLADA algorithm in Section \ref{sec:A} with a trilogy pattern. First, a basic two-layer augmented Lagrangian-based algorithm (ELL) is introduced and its convergence is discussed. Then the approximate solution of equality-constrained NLP problems and the Anderson acceleration scheme are incorporated to form the ELLA and ELLADA algorithms. The implementation of the ELLADA algorithm on the distributed optimization problem involved in distributed nonlinear MPC is shown in Section \ref{sec:MPC}, and the case study is examined in Section \ref{sec:tanks}. Conclusions and discussions are given in Section \ref{sec:conclusion}.

\section{ADMM and Its Modifications}\label{sec:ADMM}
\subsection{ADMM}
\par The alterating direction method of multipliers is the most commonly used algorithm for distributed optimization under linear equality constraints \cite{boyd2011distributed}. Specifically, consider the following problem
\begin{equation}\label{eq:optimization.problem}
\begin{small}\begin{aligned}
\min_{x, \bar{x}} \enskip f(x) + g(\bar{x}) \quad
\mathrm{s.t.} \enskip Ax + B\bar{x} = 0
\end{aligned}\end{small}
\end{equation}
with two blocks of variables $x$ and $\bar{x}$, where $f$ and $g$ are usually assumed to be convex. (The symbols in \eqref{eq:optimization.problem} are not related to the ones in Section \ref{sec:MPC}.) The augmented Lagrangian for such a constrained optimization problem is 
\begin{equation}
L(x,\bar{x};y) = f(x)+g(\bar{x}) + y^\top(Ax+B\bar{x}) + \frac{\rho}{2}\|Ax+B\bar{x}\|^2,
\end{equation}
in which $y$ stands for the vector of dual variables (Lagrangian multipliers) and $\rho>0$ is called the penalty parameter. According to the duality theory, the optimal solution should be determined by a saddle point of the augmented Lagrangian:
\begin{equation}
\begin{small}\begin{aligned}
\sup_{y}\min_{x,\bar{x}} \enskip L(x, \bar{x}; y).
\end{aligned}\end{small}
\end{equation}

\par The classical method of multipliers (MM) deals with this saddle point problem with an iterative procedure, where the primal variables are optimized first and then the dual variables are updated with a subgradient ascent \cite[Chapter~6]{bertsekas2016nonlinear}: 
\begin{equation}
\begin{small}\begin{aligned}
(x^{k+1}, \bar{x}^{k+1}) &= \arg\min_{x, \bar{x}} L(x, \bar{x}; y^{k}), \\
y^{k+1} &= y^{k} + \rho(Ax^{k+1} + B\bar{x}^{k+1}), \\
\end{aligned}\end{small}
\end{equation}
in which the superscript stands for the count of iterations. In a distributed context, $x$ and $\bar{x}$ usually can not be optimized simultaneously. ADMM is thus an approximation of MM that allows the optimization of $x$ and $\bar{x}$ to be performed separately, i.e.,
\begin{equation}\label{eq:ADMM}
\begin{small}\begin{aligned}
x^{k+1} &= \arg\min_{x} L(x, \bar{x}^k; y^k), \\
\bar{x}^{k+1} &= \arg\min_{\bar{x}} L(x^k, \bar{x}; y^k), \\
y^{k+1} &= y^{k} + \rho(Ax^{k+1} + B\bar{x}^{k+1}). \\
\end{aligned}\end{small}
\end{equation}

\par Since the appearance of ADMM in 1970s \cite{glowinski1975approximation, gabay1976dual}, there have been many works regarding its theoretical properties, extensions and applications. As we have mentioned in the Introduction, ADMM is known to have a linear convergence rate for convex problems. This does not change when the variables are constrained in convex sets. For example, if $x\in\mathcal{X}$, it suffices to modify the corresponding term $f(x)$ in the objective function by adding an indicator function $\mathbb{I}_\mathcal{X}(x)$ (equal to 0 if $x\in\mathcal{X}$ and $+\infty$ otherwise), which is still a convex function.

\subsection{ADMM with approximate updates}
\par Unless the objective terms $f(x)$ and $g(\bar{x})$ are of simple forms such as quadratic functions, the optimization of $x$ and $\bar{x}$ in \eqref{eq:ADMM} does not have an exact solution. Usually, iterative algorithms for nonlinear programming need to be called for the first two lines of \eqref{eq:ADMM}, and always searching for a highly accurate solution in each ADMM iteration will result in an excessive computational cost. It is thus desirable to solve the optimization subproblems in ADMM inexactly when the dual variables are yet far from the optimum, i.e., to allow $x^{k+1}$ and $\bar{x}^{k+1}$ to be chosen such that
\begin{equation}
\begin{small}\begin{aligned}
d_x^{k+1} \in \partial_x L(x^{k+1}, \bar{x}^{k}; y^k), \enskip 
d_{\bar{x}}^{k+1} \in \partial_{\bar{x}} L(x^{k+1}, \bar{x}^{k+1}; y^k), \enskip
\end{aligned}\end{small}
\end{equation} 
where $\partial_x$ and $\partial_{\bar{x}}$ represent the subgradients with respect to $x$ and $\bar{x}$, respectively, and $d_x$ and $d_{\bar{x}}$ are not exactly 0 but only converging to 0 asymptotically. For example, one can assign externally a shrinking and summable sequence of absolute errors \cite{eckstein1992douglas}:
\begin{equation}
\begin{small}\begin{aligned}
\|d_x^{k}\|\leq\epsilon_x^{k}, \enskip \|d_{\bar{x}}^{k}\|\leq\epsilon_{\bar{x}}^{k}, \enskip 
\sum_{k=1}^\infty \epsilon_x^{k} < \infty, \enskip \sum_{k=1}^\infty \epsilon_{\bar{x}}^{k} < \infty,  
\end{aligned}\end{small}
\end{equation}
or a sequence of relative errors to the errors proportional to other variations in the algorithm \cite{eckstein2017approximate, xie2017inexact}. 
\par It was shown in \cite{eckstein2017approximate} that a relative error criterion for terminating the iterations in subproblems, compared to other approximation criteria such as a summable absolute error sequence, better reduces the total number of subroutine iterations throughout the ADMM algorithm. Such a relative error criterion is a \emph{constructive} one, rendered to guarantee the decrease of a quadratic distance between the intermediate solutions $(x^{k}, \bar{x}^{k}, y^{k})$ and the optimum $(x^\ast, \bar{x}^\ast, y^\ast)$. In the context of distributed optimization problems under nonconvex constraints, since the convergence proof is established on a different basis from the quadratic distance, the construction of such a criterion must be reconsidered. We will address this issue in Subsection \ref{subsec:A2}.

\subsection{Anderson acceleration}\label{subsec:Anderson}
\par Linear convergence of the classical ADMM is essentially the result of subgradient dual update, which uses the information of only the first-order derivatives with respect to the dual variables: $\partial_y L = Ax + B\bar{x}$. The idea of creating a \emph{quadratically convergent} algorithm using only first-order derivatives originates back from Nesterov's approach of solving convex optimization problems, which performs iterations based on a linear extrapolation of the previous two iterations instead of the current solution alone \cite{nesterov1983method}. Such a \emph{momentum} method can be used to accelerate the ADMM algorithm, which can be seen as iterations over the second block of primal variables $\bar{x}$ and the dual variables $y$ \cite{goldstein2014fast}. However, such a momentum is inappropriate for nonconvex problems, since the behavior of the extrapolated point can not be well controlled by a bound on the curvature of the objective function. 

\par Therefore, we resort to a different type of technique -- Anderson acceleration, which was proposed in \cite{anderson1965iterative} first and later ``rediscovered" in the field of chemical physics \cite{pulay1980convergence}. Generally speaking, Anderson acceleration is used to solve the fixed-point iteration problem
\begin{equation}
w = h_0(w)
\end{equation} 
for some vector $w$ and non-expansive mapping $h_0$ (satisfying $\|h_0(w)-h_0(w^\prime)\| \leq \|w-w^\prime\|$ for any $w$ and $w^\prime$). Different from the simple Krasnoselskii-Mann iteration $w^{k+1} = \kappa w^k + (1-\kappa)h_0(w^k)$ ($\kappa\in(0,1)$), Anderson acceleration takes a quasi-Newton approach, which aims at a nearly quadratic convergence rate \cite{fang2009two}. Specifically\footnote{There are two different types of Anderson acceleration. Here we focus on Type I, which was found to have better performance \cite{fang2009two} and was improved in \cite{zhang2018globally}.}, in each iteration $k$, the results from the previous $m$ iterations are recalled from memory to form the matrix of secants in $w$ and $h(w) = w-h_0(w)$:
\begin{equation}
\begin{small}\begin{aligned}
& \Delta_w^k = \begin{bmatrix} \delta_w^{k-m} & \dots & \delta_w^{k-1} \end{bmatrix}, \\
& \delta_w^{k^\prime} = w^{k^\prime+1} - w^{k^\prime}, \enskip k^\prime = k-m,\dots,k-1; \\
& \Delta_h^k = \begin{bmatrix} \delta_h^{k-m} & \dots & \delta_h^{k-1} \end{bmatrix}, \\
& \delta_h^{k^\prime} = h(w^{k^\prime+1}) - h(w^{k^\prime}), \enskip k^\prime = k-m,\dots,k-1. \\
\end{aligned}\end{small}
\end{equation}
An estimated Jacobian is given by
\begin{equation}\label{eq:Jacobian.estimate}
H_k = I+(\Delta_h^k - \Delta_w^k)(\Delta_w^{k\top} \Delta_w^k)^{-1} \Delta_w^{k\top},
\end{equation}
or
\begin{equation}\label{eq:inverse.Jacobian.estimate}
H_k^{-1} = I+(\Delta_w^k - \Delta_h^k)(\Delta_w^{k\top} \Delta_h^k)^{-1} \Delta_w^{k\top},
\end{equation}
which minimizes the Frobenius norm of $B_k-I$ subject to $B_k\Delta_w^k = \Delta_h^k$. Then the quasi-Newton iteration $w^{k+1} = w^k - H_k^{-1}h^k$ leads to a weighted sum of the previous $m$ function values:
\begin{equation}\label{eq:Anderson}
w^{k+1} = \sum_{m^\prime=0}^m \alpha_{m^\prime}^k h_0(x^{k-m+m^\prime})
\end{equation}
where the weights $\{\alpha_{m^\prime}^k\}_{m^\prime=0}^m$ are specified by
\begin{equation}
\begin{small}\begin{aligned}
\alpha_{m^\prime}^k = 
\begin{cases}
s_0^k, & m^\prime = 0 \\
s_{m^\prime}^k - s_{m^\prime -1}^k, & m^\prime = 1,\dots,m-1 \\
1-s_{m-1}^k, & m^\prime = m \\
\end{cases},
\end{aligned}\end{small}
\end{equation}
with $s_{m^\prime}^k$ being the $m^\prime$-th component $s^k$:
\begin{equation}
s^k = (\Delta_w^{k\top} \Delta_h^k)^{-1} \Delta_w^{k\top}h^k.
\end{equation}

\par Anderson acceleration \eqref{eq:Anderson} may not always be convergent, although local convergence was studied in some special cases \cite{toth2015convergence}. Recently, a globally convergent modification of Anderson acceleration was proposed in \cite{zhang2018globally}, where regularization, restarting, and safeguarding measures are taken to ensure the well-conditioning of the $\Delta_w^k$ matrix, boundedness of the inverse Jacobian estimate \eqref{eq:inverse.Jacobian.estimate}, and acceleration only in a safety region, respectively.

\par The relevance of Anderson acceleration to ADMM lies in that the ADMM algorithm \eqref{eq:ADMM} can be seen as fixed-point iterations $(\bar{x}^k, y^k) \rightarrow (\bar{x}^{k+1}, y^{k+1})$, $k=0,1,2,\dots$ \cite{zhang2019accelerating}, which is the same idea underlying the ADMM with Nesterov acceleration. For problems with nonconvex constraints, the iteration mapping $h$ is not necessarily non-expansive, and hence one can not directly establish the convergence of Anderson acceleration with the original techniques used in \cite{zhang2018globally}. We will address this issue in Subsection \ref{subsec:A3}.

\subsection{ADMM under nonconvex constraints}
\par The presence of nonconvexity largely increases the difficulty of distributed optimization. Most of the work in nonconvex ADMM considers problems with nonconvex objective function with bounded Hessian eigenvalues or the Kurdyka-{\L}ojasiewicz property assumptions, under which a convergence rate of $\mathcal{O}(1/\sqrt{k})$ (slower than that of convex ADMM, $\mathcal{O}(1/k)$) was established \cite{li2015global, hong2016convergence, wang2019global}. However, for many distributed optimization problems, e.g., the distributed MPC of nonlinear processes, there exist nonconvex constraints on the variables, which is intrinsically non-equivalent to the problems with nonconvex objective functions. For our problem of interest, the relevant works are scarce. 

\par Here we introduce the algorithm of \cite{sun2019two} for \eqref{eq:optimization.problem} under nonconvex constraints $x\in\mathcal{X}$ and $\bar{x}\in\bar{\mathcal{X}}$, reformulated with slack variables $z$:
\begin{equation}\label{eq:Sun.problem}
\begin{small}\begin{aligned}
\min_{x, \bar{x}, z} \enskip & f(x) + g(\bar{x}) \\
\mathrm{s.t.} \enskip & Ax + B\bar{x} + z = 0, \enskip z = 0, \enskip x \in \mathcal{X}, \enskip \bar{x}\in \bar{\mathcal{X}}.
\end{aligned}\end{small}
\end{equation}
The augmented Lagrangian is now written as
\begin{equation}\label{eq:AL}
\begin{small}\begin{aligned}
& L(x,\bar{x},z;y,\lambda,\rho,\beta) = f(x)+g(\bar{x}) + \mathbb{I}_\mathcal{X}(x) + \mathbb{I}_{\bar{\mathcal{X}}}(\bar{x})  \\
& + y^\top (Ax+B\bar{x}+z) + \frac{\rho}{2} \|Ax+B\bar{x}+z\|^2 + \lambda^\top z + \frac{\beta}{2}\|z\|^2.
\end{aligned}\end{small}
\end{equation}
The algorithm is a two-layer one, where each outer iteration (indexed by $k$) contains a series of inner iterations (indexed by $r$). In the inner iterations, the classical ADMM algorithm is used to update $x$, $\bar{x}$, $z$ and $y$ in sequence, while keeping $\lambda$ and $\beta$ unchanged: 
\vspace{-13pt}
\begin{center}
\begin{small}
\begin{align}
x^{k,r+1} &= \arg\min_{x} L(x,\bar{x}^{k,r}, z^{k,r}; y^{k,r}, \lambda^{k},\rho^{k},\beta^{k}) \nonumber \\
&= \arg\min_{x\in\mathcal{X}} f(x) + \frac{\rho^k}{2}\left\| Ax+B\bar{x}^{k,r} + z^{k,r} + \frac{y^{k,r}}{\rho^k} \right\|^2 \nonumber \\
\bar{x}^{k,r+1} &= \arg\min_{\bar{x}} L(x^{k,r+1}, \bar{x}, z^{k,r}; y^{k,r}, \lambda^{k},\rho^{k},\beta^{k}) \\
&= \arg\min_{\bar{x}\in\bar{\mathcal{X}}} g(\bar{x}) + \frac{\rho^k}{2}\left\| Ax^{k,r+1}+B\bar{x} + z^{k,r} + \frac{y^{k,r}}{\rho^k} \right\|^2 \nonumber \\
z^{k,r+1} &= \arg\min_{z} L(x^{k,r+1}, \bar{x}^{k,r+1}, z; y^{k,r}, \lambda^{k},\rho^{k},\beta^{k}) \nonumber \\
&= -\frac{\rho^k}{\rho^k+\beta^k}\left( Ax^{k,r+1} + B\bar{x}^{k,r+1} + \frac{y^{k,r}}{\rho^k} \right) - \frac{1}{\rho^k+\beta^k}\lambda^k \nonumber \\
y^{k,r+1} &= y^{k,r} + \rho^k(Ax^{k,r+1} + B\bar{x}^{k,r+1} + z^{k,r+1}) \nonumber
\end{align}
\end{small}
\end{center}
Under mild assumptions, in the presence of slack variables $z$, it was proved \cite{sun2019two} that if one chooses $\rho^k = 2\beta^k$, then the inner iterations converge to the set of stationary points $(x^k, \bar{x}^k, z^k, y^k)$ of the relaxed problem
\begin{equation}\label{eq:relaxed.problem}
\begin{small}\begin{aligned}
\min_{x,\bar{x}, z}\enskip & f(x)+g(\bar{x}) + \lambda^{k\top} z + \frac{\beta^k}{2}\|z\|^2 \\
\mathrm{s.t.} \enskip & Ax+ B\bar{x} + z = 0, \enskip x\in \mathcal{X}, \enskip \bar{x}\in \bar{\mathcal{X}}.
\end{aligned}\end{small}
\end{equation}
\par Then in the outer iterations, the dual variables $\lambda^k$ are updated. To enforce the convergence of the slack variables to zero, the corresponding penalty $\beta^k$ is amplified by a ratio $\gamma > 1$ if the returned $z^k$ from the inner iterations does not decay enough from the previous outer iteration $z^{k-1}$ ($\|z^k\|>\omega\|z^{k-1}\|$, $\omega\in(0,1)$). The outer iteration is written as
\begin{equation}\label{eq:outer.dual.update}
\begin{small}\begin{aligned}
\lambda^{k+1} &= \Pi_{[\underline{\lambda}, \overline{\lambda}]}(\lambda^k + \beta^k z^k) \\
\beta^{k+1} &= 
\begin{cases}
\gamma\beta^k, &\|z^k\|>\omega\|z^{k-1}\| \\
\beta^k, &\|z^k\|\leq\omega\|z^{k-1}\|
\end{cases}
\end{aligned}\end{small}
\end{equation}
in which the projection $\Pi$ onto a predefined compact hypercube $[\underline{\lambda}, \overline{\lambda}]$ is used to guarantee the boundedness of the dual variables and hence the augmented Lagrangian $L$. If the augmented Lagrangian $L$ remains bounded despite the increase of the penalty parameters $\rho^k$ and $\beta^k$, the algorithm converges to a stationary point of the original problem \eqref{eq:optimization.problem}. The iterative complexity of such an algorithm to reach an $\epsilon$-approximate stationary point is $\mathcal{O}(\epsilon^{-4}\ln(\epsilon^{-1}))$. 
\par In the next section, building on the algorithm of \cite{sun2019two} that guarantees the convergence of distributed optimization under nonconvex constraints, we propose a new algorithm that integrates into it the ideas of approximate ADMM and Anderson acceleration, aiming at improving the computational efficiency.

\section{Proposed Algorithm}\label{sec:A}
\subsection{Basic algorithm and its convergence}\label{subsec:A1}
\par Consider an optimization problem in the following form:
\begin{equation}\label{eq:problem.1}
    \begin{small}\begin{aligned}
    \min_{x, \bar{x}} \enskip & f(x) + g(\bar{x}) \\
    \mathrm{s.t.} \enskip & Ax + B\bar{x} = 0 \\
    & x \in \mathcal{X} = \{x|\phi(x)\leq 0, \psi(x)=0\}, \enskip \bar{x} \in \bar{\mathcal{X}} \\
    \end{aligned}\end{small}
\end{equation}
or equivalently with slack variables
\begin{equation}\label{eq:problem.2}
    \begin{small}\begin{aligned}
    \min_{x, \bar{x}, z} \enskip & f(x) + g(\bar{x}) \\
    \mathrm{s.t.} \enskip & Ax + B\bar{x} + z = 0, \enskip z = 0, \\
    & x \in \mathcal{X} = \{x|\phi(x)\leq 0, \psi(x)=0\}, \enskip \bar{x} \in \bar{\mathcal{X}}.
    \end{aligned}\end{small}
\end{equation}
We make the following assumptions.
\begin{assumption}\label{assum:1}
Assume that $f$ is lower bounded, i.e., there exists $\underline{f}$ such that $f(x)\geq \underline{f}$ for any $x\in\mathcal{X}$. 
\end{assumption}
\begin{assumption}\label{assum:2}
Function $g$ is convex and is lower bounded.
\end{assumption}
Our basic algorithm (Algorithm \ref{alg:A1}) for \eqref{eq:problem.2} is slightly modified from the procedure of \cite{sun2019two}, which considered the case where $g(x)=0$ and $\bar{\mathcal{X}}$ is a hypercube. The algorithm uses an inner loop of ADMM iterations and an outer loop of MM with possibly amplifying penalty parameters. The inner iterations are terminated when the following criterion is met
\begin{equation}\label{eq:outer.termination}
    \begin{small}\begin{aligned}
    \epsilon_1^k & \geq \epsilon_1^{k,r} := \|\rho^k A^\top (B\bar{x}^{k,r+1} + z^{k,r+1} - B\bar{x}^{k,r} - z^{k,r}) \|, \\
    \epsilon_2^k & \geq \epsilon_2^{k,r} := \|\rho^k B^\top (z^{k,r+1} - z^{k,r}) \|, \\
    \epsilon_3^k & \geq \epsilon_3^{k,r} := \|Ax^{k,r+1} + B\bar{x}^{k,r+1} + z^{k,r+1} \|. \\
    \end{aligned}\end{small}
\end{equation}

\SetKw{KwInit}{Initialization:}
\SetKw{KwSet}{Set:}
\begin{algorithm}[!t]
\begin{footnotesize}
 \SetAlgoLined
 \KwSet{Bound of dual variables $[\underline{\lambda}, \overline{\lambda}]$, shrinking ratio of slack variables $\omega\in[0, 1)$, amplifying ratio of penalty parameter $\gamma>1$, diminishing outer iteration tolerances $\{\epsilon_1^k, \epsilon_2^k, \epsilon_3^k\}_{k=1}^\infty \downarrow 0$, terminating tolerances ${\epsilon}_1$, ${\epsilon}_2$, ${\epsilon}_3>0$}\; 
 \KwInit{Starting points $x^0$, $\bar{x}^0$, $z^0$, dual variable and bounds $\lambda^1 \in [\underline{\lambda}, \overline{\lambda}]$, penalty parameter $\beta^1>0$}\;
 outer iteration count $k \leftarrow 0$\;
 \While{stationarity criterion \eqref{eq:stationary.point} is not met}{
  $\rho^k = 2\beta^k$\;
  inner iteration count $r \leftarrow 0$\;
  \KwInit{$x^{k,0}$, $\bar{x}^{k,0}$, $z^{k,0}$, $y^{k,0}$ satisfying $\lambda^k + \beta^k z^{k,0} + y^{k,0} = 0$}\;
  \While{stopping criterion \eqref{eq:outer.termination} is not met}{
  $x^{k, r+1} = \arg\min_{x\in\mathcal{X}} f(x) + \frac{\rho^k}{2}\left\| Ax+B\bar{x}^{k,r} + z^{k,r} + \frac{y^{k,r}}{\rho^k} \right\|^2$\;
  $\bar{x}^{k,r+1} = \arg\min_{\bar{x}\in\bar{\mathcal{X}}} g(\bar{x}) + \frac{\rho^k}{2}\left\| Ax^{k,r+1}+B\bar{x} + z^{k,r} + \frac{y^{k,r}}{\rho^k} \right\|^2$\;
  $z^{k,r+1} = -\frac{\rho^k}{\rho^k+\beta^k}\left( Ax^{k,r+1} + B\bar{x}^{k,r+1} + \frac{y^{k,r}}{\rho^k} \right) - \frac{1}{\rho^k+\beta^k}\lambda^k$\;
  $y^{k,r+1} = y^{k,r} + \rho^k(Ax^{k,r+1} + B\bar{x}^{k,r+1} + z^{k,r+1})$\;
  $r \leftarrow r+1$\;
  }
  $(x^{k+1}, \bar{x}^{k+1}, z^{k+1}, y^{k+1}) \leftarrow (x^{k,r}, \bar{x}^{k,r}, z^{k,r}, y^{k,r})$\;
  $\lambda^{k+1} = \Pi_{[\underline{\lambda}, \overline{\lambda}]}(\lambda^k + \beta^kz^k)$\;
  \eIf{$\|z^{k+1}\| > \omega\|z^k\|$}
  {$\beta^{k+1} \leftarrow \gamma\beta^k$\;}
  {$\beta^{k+1} \leftarrow \beta^k$\;}
  $k \leftarrow k+1$\;
 }
 \caption{Basic algorithm (ELL).}\label{alg:A1}
\end{footnotesize}
\end{algorithm}

\par The proof uses the augmented Lagrangian \eqref{eq:AL} as a decreasing Lyapunov function throughout the inner iterations \cite{hong2016convergence, hong2017linear}, which gives the convergence of the inner iterations. 
\begin{lemma}[Descent of the augmented Lagrangian]\label{lemma:descent}
Suppose that Assumptions \ref{assum:1} and \ref{assum:2} hold. When $\rho^k = 2\beta^k$, it holds that
\begin{equation}\label{eq:Lyapunov}
\begin{small}\begin{aligned}
     & L(x^{k,r+1}, \bar{x}^{k,r+1}, z^{k,r+1}, y^{k,r+1}) \leq L(x^{k,r}, \bar{x}^{k,r}, z^{k,r}, y^{k,r}) \\
     & - \beta^k\|B\bar{x}^{k,r+1} - B\bar{x}^{k,r}\|^2 - \frac{\beta^k}{2}\|z^{k,r+1} - z^{k,r}\|^2
\end{aligned}\end{small}
\end{equation}
for $r=0, 1, 2, \dots$\footnote{For simplicity we did not write the last three entries $\lambda^k$, $\rho^k$, $\beta^k$ that do not change during inner iterations in the augmented Lagrangian.}, and hence the augmented Lagrangian nonincreasingly converges to a limit $\underline{L}_k$. 
\end{lemma}
\begin{corollary}[Convergence of inner iterations]\label{corollary:descent}
Suppose that Assumptions \ref{assum:1} and \ref{assum:2} hold. As $r \rightarrow \infty$, $B\bar{x}^{k,r+1} - B\bar{x}^{k,r}\rightarrow 0$, $z^{k,r+1}-z^{k,r} \rightarrow 0$, and $Ax^{k,r} + B\bar{x}^{k,r} + z^{k,r} \rightarrow 0$. Hence the inner iterations are terminated at a finite $r$ when \eqref{eq:outer.termination} is met and the point $(x^{k,r+1}, \bar{x}^{k,r+1}, z^{k,r+1})$ the following conditions
\begin{equation}\label{eq:outer.termination.approximation}
    \begin{small}\begin{aligned}
    d_1^k & \in \partial f(x^{k,r+1}) + \mathcal{N}_\mathcal{X}(x^{k,r+1}) + A^\top y^{k,r+1} \\
    d_2^k & \in \partial g(\bar{x}^{k,r+1}) + \mathcal{N}_{\bar{\mathcal{X}}}(\bar{x}^{k,r+1}) + B^\top y^{k,r+1} \\
    0 & = \lambda^k + \beta^k z^{k,r+1} + y^{k,r+1} \\
    d_3^k & = Ax^{k,r+1} + B\bar{x}^{k,r+1} + z^{k,r+1} \\
    \end{aligned}\end{small}
\end{equation}
for some $d_1^k$, $d_2^k$ and $d_3^k$ satisfying $\|d_1^k\|\leq \epsilon_1^k$, $\|d_2^k\|\leq \epsilon_2^k$ and $\|d_3^k\|\leq \epsilon_3^k$, respectively. $\mathcal{N}_{\mathcal{X}}(x)$ ($\mathcal{N}_{\bar{\mathcal{X}}}(\bar{x})$) refers to the normal cone to the set $\mathcal{X}$ ($\bar{\mathcal{X}}$) at point $x$ ($\bar{x}$):
\begin{equation}
    \mathcal{N}_\mathcal{X}(x) = \{v| v^\top (x^\prime - x) \leq 0, \enskip \forall x^\prime\in\mathcal{X}\}.
\end{equation}
\end{corollary}
The proofs of the above lemma and corollary are given in Appendix \ref{app:A} and Appendix \ref{app:B}, respectively. It is apparent that if $\epsilon_1^k, \epsilon_2^k, \epsilon_3^k$ are all equal to 0, \eqref{eq:outer.termination.approximation} is the Karush-Kuhn-Tucker optimality condition of the relaxed problem \eqref{eq:relaxed.problem} \cite{rockafellar2000variational}. 
\par We note that although the augmented Lagrangian decreases throughout the inner iterations, the increase in the penalty parameters may cause an increase in the augmented Lagrangian across outer iterations, thus losing the guarantee of overall convergence. To establish the convergence of outer iterations, we need to make the following assumption to restrict the upper level of the augmented Lagrangian.
\begin{assumption}\label{assum:3}
The augmented Lagrangians are uniformly upper bounded at initialization of all inner iterations, i.e., there exists $\overline{L}\geq L(x^{k,0}, \bar{x}^{k,0}, z^{k,0}, y^{k,0}, \lambda^k, \rho^k, \beta^k)$ for all $k$.
\end{assumption}
The above assumption is actually a ``warm start" requirement. Suppose that we have a feasible solution $(x^0, \bar{x}^0)$ to the original problem \eqref{eq:problem.1}, then we can always choose $x^{k,0} = x^0$, $\bar{x}^{k,0} = \bar{x}^0$, $z^{k,0} = 0$, $y^{k,0} = -\lambda^k$ to guarantee an $\overline{L} = f(x^0) + g(\bar{x}^0)$. 

\begin{lemma}[Convergence of outer iterations]\label{lemma:convergence.1}
Suppose that Assumptions \ref{assum:1}, \ref{assum:2} and \ref{assum:3} hold. Then for any $\epsilon_1$, $\epsilon_2$, and $\epsilon_3>0$, within a finite number of outer iterations $k$, Algorithm \ref{alg:A1} finds an approximate stationary point $(x^{k+1}, \bar{x}^{k+1}, z^{k+1}, y^{k+1})$ of \eqref{eq:problem.1}, satisfying
\begin{equation}\label{eq:stationary.point}
    \begin{small}\begin{aligned}
    d_1 & \in \partial f(x^{k+1}) + \mathcal{N}_\mathcal{X}(x^{k+1}) + A^\top y^{k+1} \\
    d_2 & \in \partial g(\bar{x}^{k+1}) + \mathcal{N}_{\bar{\mathcal{X}}}(\bar{x}^{k+1}) + B^\top y^{k+1} \\
    d_3 & = Ax^{k+1} + B\bar{x}^{k+1} \\
    \end{aligned}\end{small}
\end{equation}
for some $d_1$, $d_2$, $d_3$ satisfying $\|d_j\|\leq\epsilon_j$, $j=1,2,3$. 
\end{lemma}
See Appendix \ref{app:C} for a proof. In addition to the convergence, we can also establish a theoretical complexity. Previously in \cite{sun2019two}, it was shown that to reach an $\epsilon$-approximate stationary point satisfying \eqref{eq:stationary.point} with $\epsilon_1, \epsilon_2, \epsilon_3 = \epsilon > 0$, the total number of inner iterations needed is of the order $\mathcal{O}(\epsilon^{-4}\ln(1/\epsilon))$. Here, we show that by appropriately choosing the way that the tolerances $(\epsilon_1^k, \epsilon_2^k, \epsilon_3^k)$ shrink, the iteration complexity can be provably reduced anywhere in $(\mathcal{O}(\epsilon^{-2}), \mathcal{O}(\epsilon^{-4})]$, for which a proof is given in Appendix \ref{app:D}. 
\begin{lemma}[Complexity of the basic algorithm]\label{lemma:complexity.1}
Suppose that Assumptions \ref{assum:1}, \ref{assum:2} and \ref{assum:3} hold. For some constant $\vartheta\in(0,\omega]$, choose $\epsilon_1^k \sim \mathcal{O}(\vartheta^k)$, $\epsilon_2^k \sim \mathcal{O}(\vartheta^k)$, and $\epsilon_3^k \sim \mathcal{O}((\vartheta/\beta)^k)$. Then each outer iteration $k$ requires $R^k \sim \mathcal{O}((\vartheta\omega)^{-2k})$ inner iterations. Hence, for the Algorithm \ref{alg:A1} to reach an $\epsilon$-approximate stationary point, the total iterations needed is $R\sim\mathcal{O}(\epsilon^{-2(1+\varsigma)})$, where $\varsigma = \log_\vartheta \omega \in (0,1]$. 
\end{lemma}

\subsection{Approximate algorithm}\label{subsec:A2}
\par We note that the basic algorithm requires complete minimization of $x$ and $\bar{x}$ in each inner iteration (Lines 9--10, Algorithm \ref{alg:A1}). However, this is neither desirable due to the computational cost, nor practical since any nonlinear programming (NLP) solver finds only a point that approximately satisfies the KKT optimality conditions, except for very simple cases. For simplicity, we assume that such a minimization oracle\footnote{We use the word ``oracle" with its typical meaning in mathematics and computer science. An oracle refers to an ad hoc numerical or computational procedure, regarded as a black box mechanism, to generate the needed results as its outputs based on some input information.}, namely an explicit mapping $G$ depending on matrix $B$, $Ax^{k,r+1} + z^{k,r} + y^{k,r}/\rho^k$, and $\rho^k$, exists for $\bar{x}$. For example, if $g(\bar{x})=0$ and $B^\top B = aI$ for some $a>0$, then $G(B,v,\rho) = -\frac{1}{2a}B^\top v$. For the $x$-minimization, however, such an oracle usually does not exist. In this subsection, we will modify Algorithm \ref{alg:A1} so as to allow approximate $x$-optimization on Line 9. 
\begin{assumption}\label{assum:4}
    The minimization of the augmented Lagrangian with respect to $\bar{x}$ (Line 10, Algorithm \ref{alg:A1}) admits a unique explicit solution
    \begin{equation}\label{eq:G.mapping}
        \bar{x}^{k,r+1} = G(B, Ax^{k,r+1}+z^{k,r}+y^{k,r}/\rho^k, \rho^k).
    \end{equation}
\end{assumption}
Let us also assume that the problem has a smoothness property as follows. 
\begin{assumption}\label{assum:5}
    Functions $f$, $\phi$ and $\psi$ are continuously differentiable, and $\mathcal{X}$ has a nonempty interior.
\end{assumption}

\par Under this smoothness assumption, the KKT condition for $x$-minimization is written as the following equalities with $\mu\geq0$ and $\nu$ representing the Lagrangian dual variables corresponding to the inequalities $\phi(x)\leq 0$ and $\psi(x)=0$, respectively
\begin{equation}
    \begin{small}\begin{aligned}
    0 &= \nabla f(x^{k,r+1}) + \rho^k A^\top (Ax^{k,r+1} + B\bar{x}^{k,r} + z^{k,r} + y^{k,r}/\rho^k) \\
    & \quad + \sum_{c=1}^{C_\phi} \mu_c \nabla \phi_c(x^{k,r+1}) + \sum_{c=1}^{C_\psi} \nu_c \nabla \psi_c(x^{k,r+1}) \\
    0 &= \mu_c\phi_c(x^{k,r+1}), \enskip c=1,\dots, C_\phi \\
    0 &= \psi_c(x^{k,r+1}), \enskip c=1,\dots, C_\psi. \\
    \end{aligned}\end{small}
\end{equation}
Line 9 of Algorithm \ref{alg:A1} is thus to solve the above equations for $x^{k,r+1}$. This can be achieved through an interior point algorithm, which employs double-layer iterations to find the solution. In the outer iteration, a barrier technique is used to convert the inequality constraints into an additional term in the objective; the optima (or stationary points) of the resulting barrier problems converge to true optima (stationary points) as the barrier parameter converges to 0. In the inner iteration, a proper search method is used to obtain the optimum of the barrier problem. \emph{Since both the interior point algorithm and the basic ADMM algorithm \ref{alg:A1} have a double-layer structure, we consider matching these two layers.}

\par Specifically, in the $k$-th outer iteration, the function $f(x)$ is appended with a barrier term $-b^k\sum_{c=1}^{C_\phi} \ln(-\phi_c(x))$ ($b^k$ is the barrier parameter, converging to 0 as $k\rightarrow\infty$). Hence a ``barrier augmented Lagrangian" can be specified as
\begin{equation}
\begin{small}
\begin{aligned}
    L_b = L - b\sum_{c=1}^{C_\phi} \ln(-\phi_c(x)).
\end{aligned}
\end{small}
\end{equation}
Based on the arguments in the previous subsection, if the $x$-optimization step returns a $x^{k,r+1}$ minimizing $L_b$ with respect to $x$, then the inner iterations result in the descent of $L_{b^k}$, which implies the satisfaction of conditions \eqref{eq:outer.termination.approximation}, with $f$ modified by the barrier function. Obviously, if Assumption \ref{assum:3} holds for $L$, then it also holds for $L_{b^k}$ when $\mathcal{X}$ has a nonempty interior. It follows that the outer iterations can find an approximate stationary point of the original problem with the decay of barrier parameters $b^k$. 
\par However, precisely finding the $x^{k,r+1}$ that minimizes $L_b$ with respect to $x$, which is an equality-constrained NLP problem, still requires an iterative search procedure \cite{wachter2005line}. By matching the inner iterations of the interior point algorithm and the inner iterations of the ADMM, we propose to perform only \emph{a proper amount of searching steps} instead of the entire equality-constrained NLP in each inner iteration, so that the solution to the equality-constrained NLP problem can be approached throughout the inner iterations. For this purpose, we assume that we have at hand a solver that can find any approximate solution of equality-constrained NLP. 
\begin{assumption}\label{assum:6}
    Assume that for any equality-constrained smooth NLP problem
    \begin{equation}
    \begin{small}\begin{aligned}
        \min_x \enskip \chi(x) \quad \mathrm{s.t.} \enskip \psi(x)=0
    \end{aligned}\end{small}
    \end{equation}
    a solver that guarantees the convergence to any approximate stationary point of the above problem with a lower objective function is available. That is, starting from any initial point $x^0$, for any tolerances $\epsilon_4,\epsilon_5>0$, within a finite number of searches the solver finds a point $(x,\nu)$ satisfying
    \begin{equation}
    \begin{small}\begin{aligned}
        d_4 &= \nabla \chi(x) + \sum_{c=1}^{C_\psi} \nu_c\nabla\psi_c(x) \\
        d_{5c} &= \psi_c(x), \enskip c = 1,\dots, C_\psi. 
    \end{aligned}\end{small}
    \end{equation}
    for some $\|d_4\|\leq\epsilon_4$, $\|d_2\|\leq\epsilon_5$, and $f(x) \leq f(x^0)$. Such an approximate solution is denoted as $F(x^0; \chi, \psi, \epsilon_4, \epsilon_5)$. 
\end{assumption}

\par The above approximate NLP solution oracle is realizable by NLP solvers where the tolerances of the KKT conditions are allowed to be specified by the user, e.g., the IPOPT solver \cite{wachter2006implementation}. Under Assumption \ref{assum:6}, the $x$-update step on Line 9 of Algorithm \ref{alg:A1} is replaced by an approximate NLP solution
\begin{equation}\label{eq:NLP.onestep.1}
\begin{small}
\begin{aligned}
    x^{k,r+1} = F(x^{k,r}; \chi^{k,r}, \psi, \epsilon_4^{k,r}, \epsilon_5^{k,r}),
\end{aligned}
\end{small}
\end{equation}
where the objective function in the current iteration is the part of barrier augmented Lagrangian $L_{b^k}$ that is related to $x$ with the indicator function $\mathbb{I}_\mathcal{X}(x)$ excluded:
\begin{equation}\label{eq:varying.objective}
\begin{small}
\begin{aligned}
    \chi^{k,r}(x) =& f(x) - b_k\sum_{c=1}^{C_\phi} \ln(-\phi_c(x)) \\
    &+ \frac{\rho^k}{2}\left\|Ax + B\bar{x}^{k,r} + z^{k,r} + y^{k,r}/\rho^k \right\|^2 
\end{aligned}
\end{small}
\end{equation}
This approximate algorithm with inexact $x$-minimization is summarized as Algorithm \ref{alg:A2}. The inner iterations are performed until $\epsilon_4^{k,r}$ and $\epsilon_5^{k,r}$ are lower than $\epsilon_4^k$ and $\epsilon_5^k$, respectively, and \eqref{eq:outer.termination} holds. The outer iterations are terminated when $\epsilon_4^k \leq \epsilon_4$, $\epsilon_5^k \leq \epsilon_5$, the barrier parameter is sufficiently small $b^k \leq \epsilon_6$, and \eqref{eq:stationary.point} holds. 

\begin{algorithm}[!t]
\begin{footnotesize}
 \SetAlgoLined
 \KwSet{Bound of dual variables $[\underline{\lambda}, \overline{\lambda}]$, shrinking ratio of slack variables $\omega\in[0, 1)$, amplifying ratio of penalty parameter $\gamma>1$, diminishing outer iteration tolerances $\{\epsilon_1^k, \epsilon_2^k, \epsilon_3^k, \epsilon_4^k, \epsilon_5^k, \epsilon_6^k\}_{k=1}^\infty \downarrow 0$, diminishing barrier parameters $\{b^k\}_{k=1}^\infty \downarrow 0$, terminating tolerances ${\epsilon}_1$, ${\epsilon}_2$, ${\epsilon}_3$, $\epsilon_4$, $\epsilon_5$, $\epsilon_6>0$}\; 
 \KwInit{Starting points $x^0$, $\bar{x}^0$, $z^0$, dual variable and bounds $\lambda^1 \in [\underline{\lambda}, \overline{\lambda}]$, penalty parameter $\beta^1>0$}\;
 outer iteration count $k \leftarrow 0$\;
 \While{$\epsilon_4^{k}\geq \epsilon_4$ or $\epsilon_5^{k} \geq \epsilon_5$ or $b^k \geq \epsilon_6$ or stationarity criterion \eqref{eq:stationary.point} is not met}{
  \KwSet{Diminishing tolerances $\{\epsilon_4^{k,r}, \epsilon_5^{k,r}\}_{r=1}^\infty \downarrow 0$}\;
  let $\rho^k = 2\beta^k$\;
  inner iteration count $r \leftarrow 0$\;
  \KwInit{$x^{k,0}$, $\bar{x}^{k,0}$, $z^{k,0}$, $y^{k,0}$ satisfying $\lambda^k + \beta^k z^{k,0} + y^{k,0} = 0$}\;
  \While{$\epsilon_4^{k,r}\geq \epsilon_4^k$ or $\epsilon_5^{k,r} \geq \epsilon_5^k$ or stopping criterion \eqref{eq:outer.termination} is not met}{
  $x^{k,r+1} = F(x^{k,r}; \chi^{k,r}, \psi, \epsilon_4^{k,r}, \epsilon_5^{k,r})$, where $\chi^{k,r}$ is given by \eqref{eq:varying.objective}\;
  $\bar{x}^{r+1} = G(B, Ax^{r+1} + z^{k,r} + y^{k,r}/\rho^k, \rho^k)$, where $G$ is given by \eqref{eq:G.mapping}\;
  $z^{k,r+1} = -\frac{\rho^k}{\rho^k+\beta^k}\left( Ax^{k,r+1} + B\bar{x}^{k,r+1} + \frac{y^{k,r}}{\rho^k} \right) - \frac{1}{\rho^k+\beta^k}\lambda^k$\;
  $y^{k,r+1} = y^{k,r} + \rho^k(Ax^{k,r+1} + B\bar{x}^{k,r+1} + z^{k,r+1})$\;
  $r \leftarrow r+1$\;
  }
  $(x^{k+1}, \bar{x}^{k+1}, z^{k+1}, y^{k+1}) \leftarrow (x^{k,r}, \bar{x}^{k,r}, z^{k,r}, y^{k,r})$\;
  $\lambda^{k+1} = \Pi_{[\underline{\lambda}, \overline{\lambda}]}(\lambda^k + \beta^kz^k)$\;
  \eIf{$\|z^{k+1}\| > \omega\|z^k\|$}
  {$\beta^{k+1} \leftarrow \gamma\beta^k$\;}
  {$\beta^{k+1} \leftarrow \beta^k$\;}
  $k \leftarrow k+1$\;
 }
 \caption{Approximate algorithm (ELLA).}\label{alg:A2}
\end{footnotesize}
\end{algorithm}

\begin{lemma}[Convergence of the approximate algorithm]\label{lemma:convergence.2}
Suppose that Assumptions \ref{assum:1}--\ref{assum:6} hold. For any outer iteration $k$, given any positive tolerances $\{\epsilon_1^k, \dots, \epsilon_5^k\}$, within a finite number of inner iterations $r$, the obtained solution satisfies
\begin{equation}\label{eq:convergence.2.1}
    \begin{small}\begin{aligned}
    d_1^k + d_4^k & = \nabla f(x^{k,r+1}) + \sum_{c=1}^{C_\phi} \mu_c^{k,r+1}\nabla\phi_c(x^{k,r+1}) \\
    & \quad + \sum_{c=1}^{C_\psi} \nu_c^{k,r+1}\nabla\psi_c(x^{k,r+1}) + A^\top y^{k,r+1} \\
    d_2^k & \in \partial g(\bar{x}^{k,r+1}) + \mathcal{N}_{\bar{\mathcal{X}}}(\bar{x}^{k,r+1}) + B^\top y^{k,r+1} \\
    0 & = \lambda^k + \beta^k z^{k,r+1} + y^{k,r+1} \\
    d_3^k & = Ax^{k,r+1} + B\bar{x}^{k,r+1} + z^{k,r+1}, \\
    d_5^k &= \psi(x^{k,r+1})  \\
    -b^k &= \mu_c^{k,r+1}\phi_c(x^{k,r+1}), \enskip c=1, \dots, C_\phi. 
    \end{aligned}\end{small}
\end{equation}
for some $d_1^k,\dots,d_5^k$ with $\|d_1^k\| \leq \epsilon_1^k$, \dots, $\|d_5^k\| \leq \epsilon_5^k$. Then, suppose that the outer iteration tolerances $\{\epsilon_1^k,\dots, \epsilon_5^k\}$ and barrier parameters $b^k$ are diminishing with increasing $k$, given any terminating tolerances $\epsilon_1, \dots, \epsilon_6>0$, within a finite number of outer iterations, Algorithm \ref{alg:A2} finds a point $(x^{k+1}$, $\bar{x}^{k+1}$, $z^{k+1}$, $y^{k+1}$, $\mu^{k+1}$, $\nu^{k+1})$ satisfying 
\begin{equation}\label{eq:convergence.2.2}
    \begin{small}\begin{aligned}
    d_1 + d_4 &= \nabla f(x^{k+1}) + \sum_{c=1}^{C_\phi} \mu_c^{k+1} \nabla\phi_c(x^{k+1}) \\
    & \quad + \sum_{c=1}^{C_\psi} \nu_c^{k+1}\nabla\psi_c(x^{k+1}) + A^\top y^{k+1} \\
    d_2 & \in \partial g(\bar{x}^{k+1}) + \mathcal{N}_{\bar{\mathcal{X}}}(\bar{x}^{k+1}) + B^\top y^{k+1} \\
    0 & = \lambda^k + \beta^k z^{k+1} + y^{k+1} \\
    d_3 & = Ax^{k+1} + B\bar{x}^{k+1}, \\
    d_5 &= \psi(x^{k+1}) \\
    -d_6 &= \mu_c^{k+1}\phi_c(x^{k+1}), \enskip c=1, \dots, C_\phi. 
    \end{aligned}\end{small}
\end{equation}
for some $d_1,\dots,d_6$ with $\|d_j\| \leq \epsilon_j$, $j = 1,\dots, 5$, $d_6\in(0,\epsilon_6]$.
\end{lemma}
The proof is self-evident following the techniques in the Proofs of Lemma \ref{lemma:descent}, Corollary \ref{corollary:descent} and Lemma \ref{lemma:convergence.1} given in Appendix \ref{app:A} to Appendix \ref{app:C}. The conditions \eqref{eq:convergence.2.1} indicate an $(\epsilon_1^k, \dots, \epsilon_5^k)$-approximate stationary point to the relaxed barrier problem 
\begin{equation}\label{eq:relax.barrier.problem}
    \begin{small}\begin{aligned}
    \min_{x,\bar{x},z} \enskip & f(x) + g(\bar{x}) - b^k\sum_{c=1}^{C_\phi} \ln(-\phi_c(x)) \\
    \mathrm{s.t.} \enskip & Ax + B\bar{x} + z = 0, \enskip \psi(x)=0, \enskip \bar{x} \in \bar{\mathcal{X}} \\
    \end{aligned}\end{small}
\end{equation}
and the condition \eqref{eq:convergence.2.2} gives an $(\epsilon_1, \dots, \epsilon_6)$-approximate stationary point to the original problem \eqref{eq:problem.1}.

\subsection{Accelerated algorithm}\label{subsec:A3}
\par The key factor restricting the rate of convergence is the $y$-update, which is not a full or approximate maximization but only one step of subgradient ascent. As was proved in Lemma \ref{lemma:complexity.1}, such a subgradient ascent approach for nonconvex problems leads to a number of inner iterations proportional to the inverse squared error. Here, by modifying the Anderson acceleration scheme in \cite{zhang2018globally}, we propose an accelerated algorithm. Let us make the following assumption regarding our choice of tolerances $\epsilon_4^{k,r}$ and $\epsilon_5^{k,r}$.
\begin{assumption}\label{assum:7}
    Suppose that we choose a continuous and strictly monotonically increasing function $\pi: [0,\infty) \rightarrow [0,\infty)$ with $\pi(0) = 0$ such that $\epsilon_5^{k,r} = \pi(\epsilon_4^{k,r})$, and choose $\epsilon_4^{k,r+1}$ proportional to $\|\rho^k A^\top (B\bar{x}^{k,r+1} - B\bar{x}^{k,r} + z^{k,r+1} - z^{k,r})\|$ when such a value is strictly smaller than the previous tolerance $\epsilon_4^{k,r}$ but not smaller the ultimate one $\epsilon_4^k$. 
\end{assumption}
The choice of function $\pi$ to relate the stationarity tolerance and equality tolerance in NLP subroutine is aimed at balancing the effort to reduce both errors. The choice of $\epsilon_4^{k,r+1}$ is based on the following rationale. After the $r$-th inner iteration, the obtained solution $x^{k,r+1}$ satisfies the approximate stationarity condition
\begin{equation}\label{eq:barrier.inner.stationary}
    \begin{small}\begin{aligned}
    & d_4^{k,r+1} = \nabla f(x^{k,r+1}) + \sum_{c=1}^{C_\phi} \mu_{c}^{k,r+1} \nabla\phi_c(x^{k,r+1}) + \sum_{c=1}^{C_\psi} \nu_c^{k,r+1} \cdot \\
    & \nabla\psi_c(x^{k,r+1}) 
    + \rho^k A^\top \left( Ax^{k,r+1} + B\bar{x}^{k,r} + z^{k,r} + y^{k,r}/\rho^k \right)
    \end{aligned}\end{small}
\end{equation}
for some $d_4^{k,r+1}$ with a modulus not exceeding $\epsilon_4^{k,r}$, $\mu^{k,r+1}$ satisfying $\mu_c^{k,r+1}\phi_c(x^{k,r+1}) = -b^k$, $c=1,\dots,C_\phi$. Using the formula for $y$-update (Line 13, Algorithm \ref{alg:A2}), we rearrange the above equation to obtain
\begin{equation}\label{eq:barrier.outer.stationary}
    \begin{small}\begin{aligned}
    & d_4^{k,r+1} + \rho^k A^\top (B\bar{x}^{k,r+1} - B\bar{x}^{k,r} + z^{k,r+1} - z^{k,r}) = \nabla f(x^{k,r+1}) \\
    & + \sum_{c=1}^{C_\phi} \mu_{c}^{k,r+1} \nabla\phi_c(x^{k,r+1}) + \sum_{c=1}^{C_\psi} \nu_c^{k,r+1} \nabla\psi_c(x^{k,r+1}) 
     + A^\top y^{k,r+1}
    \end{aligned}\end{small}
\end{equation}
Hence after the update of $\bar{x}$, $z$ and $y$ variables, the violation of the stationarity condition is bounded by $\epsilon_4^{k,r} + \|\rho^k A^\top (B\bar{x}^{k,r+1} - B\bar{x}^{k,r} + z^{k,r+1} - z^{k,r})\|$. Therefore, $\epsilon_4^{k,r}$ should be balanced with the second term, which, however, is realizable only after the $\bar{x}$- and $z$-updates after the $x$-update and hence assigned to $\epsilon_4^{k,r+1}$. 

\par We note from Algorithm \ref{alg:A2} that under Assumption \ref{assum:7}, each inner iteration $r$ is a mapping from $(x^{k,r}, \bar{x}^{k,r}$, $z^{k,r+1}$, $y^{k,r+1}$, $\epsilon_4^{k,r})$ to $(x^{k,r+1}$, $\bar{x}^{k,r+1}$, $z^{k,r+1}$, $y^{k,r+1}$, $\epsilon_4^{k,r+1})$. In fact, despite the dependence of the latter variables on $x^{k,r}$ and $\epsilon_4^{k,r}$, such dependence can be ignored in the sense that the descent of the barrier augmented Lagrangian $L_{b^k}$ will always guide the sequence of intermediate solutions towards the set of $\epsilon_4^k$-approximate stationary points of the relaxed barrier problem \eqref{eq:relax.barrier.problem}. Using Lemma \ref{lemma:descent} with the augmented Lagrangian substituted by the barrier augmented Lagrangian, it immediately follows that under the approximate algorithm, the sequence $\{(\bar{x}^{k,r}, z^{k,r})\}_{r=1}^\infty$ will converge to a fixed point, and the convergence of $\{y^{k,r}\}$ accompanies the convergence of $\{z^{k,r}\}$ due to \eqref{eq:A1.3.4}. It is thus clear that we may resort to Anderson acceleration introduced in Subsection \ref{subsec:Anderson} by denoting $w = (\bar{x}, z)$, the iteration as a mapping $h_0$, and $h(w) = w - h_0(w)$, and collecting at the $r$-th inner iteration the following multi-secant information about the previous $m$ inner iterations:
\begin{equation}
    \begin{small}\begin{aligned}
    \Delta_w^{k,r} = [\delta_w^{k,r-m} \enskip\dots\enskip \delta_w^{k,r-1}], \enskip
    \Delta_h^{k,r} = [\delta_h^{k,r-m} \enskip\dots\enskip \delta_h^{k,r-1}],
    \end{aligned}\end{small}
\end{equation}
where $\delta_h^{r-m^\prime} = w^{k,r-m^\prime+1} - w^{k,r-m^\prime}$ and $\delta_h^{r-m^\prime} = h(w^{k,r-m^\prime+1}) - h(w^{k,r-m^\prime})$, $m^\prime = m-1,\dots, 0$. 

\par However, the possibility that $\Delta_w^k$ may not be of full rank and $H_k$ may be singular requires certain modifications to the original accelration scheme. The following technique was used in \cite{zhang2018globally}. First, it can be shown that the matrix $H$ defined in \eqref{eq:Jacobian.estimate} can be constructed in an inductive way, starting from $H_{k,r}^0 = I$, by rank-one updates
\begin{equation}\label{eq:Anderson.H.matrix.original}
\begin{small}\begin{aligned}
    H_{k,r}^{m^\prime+1} = H_{k,r}^{m^\prime} + \frac{(\delta_h^{k,r-m+m^\prime} - H_{k,r}^{m^\prime} \delta_w^{k,r-m+m^\prime})  (\hat{\delta}_w^{k,r-m+m^\prime})^\top}{(\hat{\delta}_w^{k,r-m+m^\prime})^\top \delta_w^{k,r-m+m^\prime}}
\end{aligned}\end{small}
\end{equation}
for $m^\prime = 0, \dots, m-1$ with $H_{k,r}^m = H_{k,r}$, where $\hat{\delta}_w^{k,r-m}, \dots, \hat{\delta}_w^{k,r-1}$ are obtained from $\delta_w^{k,r-m}, \dots, \delta_w^{k,r-1}$ through Gram-Schmidt orthogonalization. To ensure the invertibility of $H_{k,r}$, the $\delta_h$ vector in \eqref{eq:Anderson.H.matrix.original} is perturbed to
\begin{equation}\label{eq:theta.perturbation}
\begin{small}\begin{aligned}
    \tilde{\delta}_h^{k,r-m+m^\prime} = (1-\theta_{k,r}^{m^\prime})  \delta_h^{k,r-m+m^\prime} + \theta_{k,r}^{m^\prime} \delta_w^{k,r-m+m^\prime},
\end{aligned}\end{small}
\end{equation}
where the perturbation magnitude $\theta_{k,r}^{m^\prime}$ is determined by 
\begin{equation}
\begin{small}\begin{aligned}
    \theta_{k,r}^{m^\prime} = \varphi\left( \frac{(\hat{\delta}_w^{k,r-m+m^\prime})^\top (H_{k,r}^{m^\prime})^{-1} \delta_h^{k,r-m+m^\prime}}{\|\hat{\delta}_w^{k,r-m+m^\prime}\|^2}; \eta_\theta \right).
\end{aligned}\end{small}
\end{equation}
with regularization hyperparameter $\eta_\theta\in(0,1)$. The function $\varphi(\theta; \eta)$ is defined by
\begin{equation}
\begin{small}\begin{aligned}
    \varphi(\theta; \eta) = 
    \begin{cases} 
    (\eta\mathrm{sign}\theta-\theta)/(1-\theta) &,|\theta|\leq\eta \\
    0 &, |\theta| > \eta \\
    \end{cases}. 
\end{aligned}\end{small}
\end{equation}
Using the Sherman-Morrison formula for inverting the rank-one update, $H_{k,r}^{-1}$ can be induced from $(H_{k,r}^0)^{-1} = I$ according to
\begin{equation}\label{eq:Sherman.Morrison}
\begin{small}\begin{aligned}
& (H_{k,r}^{m^\prime+1})^{-1} = (H_{k,r}^{m^\prime})^{-1} + \\
& \frac{\left( \delta_w^{k,r-m+m^\prime} - (H_{k,r}^{m^\prime})^{-1} \tilde{\delta}_h^{k,r-m+m^\prime} \right) (\hat{\delta}_w^{k,r-m+m^\prime})^\top (H_{k,r}^{m^\prime})^{-1}}{(\hat{\delta}_w^{k,r-m+m^\prime})^\top (H_{k,r}^{m^\prime})^{-1} \tilde{\delta}_h^{k,r-m+m^\prime}}. \\
\end{aligned}\end{small}
\end{equation}

\par To avoid the rank deficiency $\Delta_w$, a restart checking strategy is used, where the memory is cleared when the Gram-Schmidt orthogonalization becomes ill conditioned ($\|\hat{\delta}_w^{k,r}\| < \eta_w\|\delta_w^{k,r}\|$ for some $\eta_w \in (0,1)$) or the memory exceeds a maximum $M$; otherwise the memory is allowed to grow. Hence the Anderson acceleration is well-conditioned. 

\begin{lemma}[Well-conditioning of Anderson acceleration, \cite{zhang2018globally}]\label{lemma:Zhang.Junzi}
Using the regularization and restart checking techniques, it is guaranteed that 
\begin{equation}
\begin{small}\begin{aligned}
    \|H_{k,r}^{-1}\|_2 \leq \theta^{-M}\left[ 3(1+\theta+\eta_w)^M \eta_w^{-N} - 2 \right]^{N-1} < +\infty
\end{aligned}\end{small}
\end{equation}
where $M$ is the maximum number of steps in the memory and $N$ is the dimension of $w$. 
\end{lemma}

\par A well-conditioned Anderson acceleration is not yet sufficient to guarantee the convergence. Hence we employ a safeguarding technique modified from \cite{zhang2018globally} which aims at suppressing a too large increase in the barrier augmented Lagrangian by rejecting such acceleration steps. When Anderson acceleration suggests an update from $w=(\bar{x}, z)$ to $\tilde{w}=(\tilde{\bar{x}}, \tilde{z})$ under the current value of $Ax$, the resulting Lagrangian increase is calculated as
\begin{equation}\label{eq:Lagrangian.increase}
\begin{small}\begin{aligned}
    \tilde{L}^k(w, \tilde{w}; Ax) =& g(\tilde{\bar{x}}) - g(\bar{x}) + \lambda^{k\top} (\tilde{z}-z) + \frac{\beta^k}{2} (\|\tilde{z}\|^2-\|z\|^2) \\
    &+ \tilde{y}^\top (Ax + B\tilde{\bar{x}} + \tilde{z}) - y^\top (Ax + B\bar{x} + z) \\
    &+ \frac{\rho^k}{2}\left( \|Ax + B\tilde{\bar{x}} + \tilde{z}\|^2  - \|Ax + B\bar{x} + \tilde{z}\|^2 \right)
\end{aligned}\end{small}
\end{equation}
where $y$ and $\tilde{y}$ are calculated by
\begin{equation}
\begin{small}\begin{aligned}
    y = -\lambda^k - \beta^k z, \enskip \tilde{y} = -\lambda^k - \beta^k \tilde{z},
\end{aligned}\end{small}
\end{equation}
which results from Line 12--13 of Algorithm \ref{alg:A2}. We require that such a change, if positive, must not exceed an upper bound:
\begin{equation}\label{eq:Lagrangian.increase.criterion}
    \tilde{L}^k(w, \tilde{w}; Ax) = \tilde{L}_0\eta_L(R_+ + 1)^{-(1+\sigma)}
\end{equation}
where $\tilde{L}_0$ is the expected Lagrangian decrease after the first non-accelerated iteration after initialization according to Lemma \ref{lemma:descent}, used as a scale for the change in the barrier augmented Lagrangian:
\begin{equation}\label{eq:Lagrangian.scale}
\begin{small}\begin{aligned}
    \tilde{L}_0 = \beta^k \|B\bar{x}^{k,1} - B\bar{x}^{k,0}\|^2 + \frac{\beta^k}{2} \|z^{k,1} - z^{k,0}\|^2,
\end{aligned}\end{small}
\end{equation}
where $\eta_L, \sigma>0$ are hyperparameters, and $R_+$ is the number of already accepted acceleration steps. With safeguarding, it can be guaranteed that the barrier augmented Lagrangian always stays bounded, since $\sum_{R_+=0}^\infty(R_+ + 1)^{-(1+\sigma)} < +\infty$. We also require that the acceleration should not lead to a drastic change in $w$:
\begin{equation}\label{eq:norm.increase.criterion}
\begin{small}\begin{aligned}
    \|\tilde{w} - w\|^2 \leq \frac{\tilde{L}_0}{\beta^k} \frac{\eta_{\tilde{w}}}{\sqrt{1+R_+}},
\end{aligned}\end{small}
\end{equation}
where $\eta_{\tilde{w}}>0$ is a hyperparameter. $1/\sqrt{1+R_+}$ reflects an expected change according to the plain ADMM iteration, which is used to suppress disproportionate large deviations due to Anderson acceleration.

\par Finally, the accelerated algorithm using the Anderson acceleration technique for fixed-point iteration of $(\bar{x}, z)$ is summarized as Algorithm \ref{alg:A3}. This is our final ELLADA algorithm, whose distributed implementation will be briefly discussed in the next subsection. With well-conditioned $H_{k,r}^{-1}$ matrix and a bounded barrier augmented Lagrangian, its convergence can now be guaranteed by the following lemma, the proof of which is given in Appendix \ref{app:E}. 
\begin{lemma}[Convergence under Anderson acceleration]\label{lemma:convergence.3}
Suppose that Assumptions \ref{assum:1}--\ref{assum:7} hold. Under regulated and safe-guarded Anderson acceleration, Algorithm \ref{alg:A3} finds within a finite number of inner iterations $r$ a point satisfying \eqref{eq:convergence.2.1}. The convergence of outer iterations to an approximate stationary point satisfying \eqref{eq:convergence.2.2} hence follows. 
\end{lemma}

\begin{algorithm*}[!t]
\begin{footnotesize}
 \SetAlgoLined
 \KwSet{Dual bounds $[\underline{\lambda}, \overline{\lambda}]$, outer iteration parameters $\omega\in[0, 1)$, $\gamma>1$, $\{\epsilon_1^k, \epsilon_2^k, \epsilon_3^k, \epsilon_4^k, \epsilon_6^k\}_{k=1}^\infty \downarrow 0$, $\{b^k\}_{k=1}^\infty \downarrow 0$, final tolerances ${\epsilon}_1$, ${\epsilon}_2$, ${\epsilon}_3$, $\epsilon_4$, $\epsilon_6>0$, function $\pi$, acceleration parameters $\theta\in (0,1)$, $\sigma>0$, $\eta_{\epsilon}>0$, $\eta_w\in(0,1)$, $\eta_L>0$, $\eta_{\tilde{w}}>0$, $M\in\mathbb{N}$}. Let $\epsilon_5 = \pi(\epsilon_4)$\;
 \KwInit{Starting points $x^0$, $\bar{x}^0$, $z^0$, $\lambda^1 \in [\underline{\lambda}, \overline{\lambda}]$, penalty parameter $\beta^1>0$, $\epsilon_5^0 = \pi(\epsilon_4^0)$}\;
 Outer iteration count $k \leftarrow 0$\;
 \While{$\epsilon_4^{k}\geq \epsilon_4$ or $\epsilon_5^{k} \geq \epsilon_5$ or $b^k \geq \epsilon_6$ or stationarity criterion \eqref{eq:stationary.point} is not met}{
  \KwSet{Initial tolerances $\epsilon_4^{k,0}$, $\epsilon_5^{k,0} = \pi(\epsilon_4^{k,0})$, penalty $\rho^k = 2\beta^k$, Jacobian estimate $H_{k,0}^{-1}=I$}\;
  Inner iteration count $r \leftarrow 0$, count of accelerated steps $R_+^k = 0$, memory length $m \leftarrow 0$\;
  \KwInit{$x^{k,0}$, $\bar{x}^{k,0}$, $z^{k,0}$, $y^{k,0}$ satisfying $\lambda^k + \beta^k z^{k,0} + y^{k,0} = 0$}\;
  \While{$\epsilon_4^{k,r}\geq \epsilon_4^k$ or $\epsilon_5^{k,r} \geq \epsilon_5^k$ or stopping criterion \eqref{eq:outer.termination} is not met}{
  $x^{k,r+1} = F(x^{k,r}; \chi^{k,r}, \psi, \epsilon_4^{k,r}, \epsilon_5^{k,r})$, where $\chi^{k,r}$ is given by \eqref{eq:varying.objective}\;
  $\bar{x}^{k,r+1} = G(B, Ax^{r+1} + z^{k,r} + y^{k,r}/\rho^k, \rho^k)$, where $G$ is given by \eqref{eq:G.mapping}\;
  $z^{k,r+1} = -\frac{\rho^k}{\rho^k+\beta^k}\left( Ax^{k,r+1} + B\bar{x}^{k,r+1} + \frac{y^{k,r}}{\rho^k} \right) - \frac{1}{\rho^k+\beta^k}\lambda^k$\;
  $y^{k,r+1} = y^{k,r} + \rho^k(Ax^{k,r+1} + B\bar{x}^{k,r+1} + z^{k,r+1})$\;
  
  $\tilde{y}^{k,r} = -\lambda^k - \beta^k \tilde{z}^{k,r}$\;
  $\tilde{x}^{k,r+1} = F(x^{k,r}; \tilde{\chi}^{k,r}, \psi, \epsilon_4^{k,r}, \epsilon_5^{k,r})$, with $\tilde{\chi}$ in \eqref{eq:varying.objective} with $\bar{x}, z, y$ replaced by $\tilde{\bar{x}}, \tilde{z}, \tilde{y}$\;
  $\tilde{\bar{x}}^{k,r+1} = G(B, A\tilde{x}^{r+1} + \tilde{z}^{k,r} + \tilde{y}^{k,r}/\rho^k, \rho^k)$\;
  $\tilde{z}^{k,r+1} = -\frac{\rho^k}{\rho^k + \beta^k} \left( A\tilde{x}^{k,r+1} + B\tilde{\bar{x}}^{k,r+1} + \frac{\tilde{y}^{k,r}}{\rho^k} \right) - \frac{1}{\rho^k+\beta^k}\lambda^k$\;
  
  \eIf{$r=0$}{
  $\tilde{w}^{k,1} \leftarrow (\bar{x}^{k,1}, z^{k,1})$, and calculate $\tilde{L}_0$ by \eqref{eq:Lagrangian.scale}\;}{
  $\delta_w^{k,r-1} = \tilde{w}^{k,r} - w^{k,r-1}$, $\delta_h^{k,r} = \tilde{w}^{k,r}-\tilde{w}^{k,r+1}-w^{k,r-1}+w^{k,r}$, $m \leftarrow m+1$\;
  $\hat{\delta}_w^{k,r-1} = \delta_w^{k,r-1} - \sum_{m^\prime = 2}^m \frac{(\hat{\delta}_w^{k,r-m^\prime})^\top \delta_w^{k,r-1}}{\|\hat{\delta}_w^{k,r - m^\prime}\|^2} \hat{\delta}_w^{k,r-m^\prime}$\;
  \lIf{$m=M+1$ or $\frac{\|\hat{\delta}_w^{k,r-1}\|}{\|\delta_w^{k,r-1}\|} < \eta_w$}{$m\leftarrow 0$, $\hat{\delta}_w^{k,r-1}\leftarrow \delta_w^{k,r-1}$, and $H_{k,r-1}^{-1}\leftarrow I$}
  
  Compute $\tilde{\delta}_h^{k,r-1}$ by \eqref{eq:theta.perturbation} with $m^\prime=m-1$ and $\theta_{k,r-1} = \varphi\left( \frac{(\hat{\delta}_w^{k,r-1})^\top H_{k,r-1}^{-1}\delta_h^{k,r-1}}{\|\hat{\delta}_w^{k,r-1}\|^2}; \theta \right)$\;
  Update $H_{k,r}^{-1} = H_{k,r-1}^{-1} + \frac{(\delta_w^{k,r-1} - H_{k,r-1}^{-1} \tilde{\delta}_h^{k,r-1}) (\hat{\delta}_w^{k,r-1})^\top H_{k,r-1}^{-1}}{(\hat{\delta}_w^{k,r-1})^\top H_{k,r-1}^{-1} \tilde{\delta}_w^{k,r-1}}$, and suggest $\tilde{w}^{k,r+1} = w^{k,r} - H_{k,r}^{-1}(w^{k,r} - w^{k,r+1})$\;
  
  \lIf{$\frac{\tilde{L}^k(w^{k,r}, \tilde{w}^{k,r+1}; Ax^{k,r})}{\tilde{L}_0\eta_L (R_++1)^{-(1+\sigma)}}\leq 1$ and $\|\tilde{w}^{k,r+1}-w^{k,r}\|^2 \leq \frac{\tilde{L}_0\eta_{\tilde{w}}}{\beta^k \sqrt{R_++1}}$}{accept the acceleration $w^{k,r+1}\leftarrow \tilde{w}^{k,r+1}$, and let $y^{k,r+1}\leftarrow -\lambda^k - \beta^k \tilde{z}^{k,r+1}$}
  }
  
  $\epsilon_4^{k,r+1} = \|\rho^k A^\top (B\bar{x}^{k,r+1}-B\bar{x}^{k,r} + z^{k,r+1} - z^{k,r})\|$, $\epsilon_5^{k,r+1} = \pi(\epsilon_4^{k,r+1})$\;
  $r \leftarrow r+1$\;
  }
  $(x^{k+1}, \bar{x}^{k+1}, z^{k+1}, y^{k+1}) \leftarrow (x^{k,r}, \bar{x}^{k,r}, z^{k,r}, y^{k,r})$\;
  Update $\lambda^{k+1} = \Pi_{[\underline{\lambda}, \overline{\lambda}]}(\lambda^k + \beta^kz^k)$ and $\beta^{k+1}$ according to \eqref{eq:outer.dual.update}\;
  $k \leftarrow k+1$\;
 }
 \caption{Accelerated algorithm (ELLADA).}\label{alg:A3}
\end{footnotesize}
\end{algorithm*}

\par Summarizing the conclusions of all the previous lemmas in this section, we have arrived at the following theorem.
\begin{theorem}
Suppose that the following assumptions hold:
\begin{enumerate}
    \item Function $f$ is lower bounded on $\mathcal{X}$;
    \item Function $g$ is convex and lower bounded on $\bar{\mathcal{X}}$;
    \item Initialization of outer iterations allows a uniform upper bound of the augmented Lagrangian, e.g., a feasible solution is known a priori;
    \item Minimization of $g(\bar{x})+\frac{\rho}{2}\|B\bar{x}+v\|^2$ with respect to $\bar{x}$ allows an oracle $G(B, v, \rho)$ returning a unique solution for any $v$ of appropriate dimension and $\rho>0$;
    \item Functions $f$, $\phi$, and $\psi$ are continuously differentiable, and the constraints $(\phi,\psi)$ are strictly feasible; 
    \item There exists a solver for equality-constrained NLP to any specified tolerances of KKT conditions.
\end{enumerate}
Then given any tolerances $\epsilon_1, \dots, \epsilon_6 > 0$, the ELLADA algorithm (Algorithm \ref{alg:A3}) gives an $(\epsilon_1, \dots, \epsilon_6)$-approximate KKT point satisfying the conditions \eqref{eq:convergence.2.2}. 
\end{theorem}
If the problem itself has intrinsically better properties to guarantee that each KKT point is a local minimum, e.g., the second-order sufficient condition \cite[\S4.3.2]{bertsekas2016nonlinear}, then the algorithm converges to a local minimum. Of course, it is well known that certifying a local minimum is itself a difficult problem.

\section{Implementation on Distributed Nonlinear MPC}\label{sec:MPC}
\par Consider a nonlinear discrete-time dynamical system
\begin{equation}\label{eq:sys}
x(t+1) = f(x(t), u(t))
\end{equation}
where $x(t)\in\mathbb{R}^n$ and $u(t)\in\mathbb{R}^m$ are the vectors of states and inputs, respectively, for $t = 0, 1, 2, \dots$, and $f: \mathbb{R}^n \times \mathbb{R}^m \rightarrow \mathbb{R}^n$. Suppose that at time $t$ we have the current states $x=x(t)$, then in MPC, the control inputs are determined by the following optimal control problem:
\begin{equation}\label{eq:OPC}
\begin{small}\begin{aligned}
\min \enskip & J=\sum_{\tau=t}^{t+T-1} \ell(\hat{x}(\tau), \hat{u}(\tau)) + \ell^\mathrm{f}(\hat{x}(t+T)) \\
\mathrm{s.t.} \enskip & \hat{x}(\tau+1) = f(\hat{x}(\tau), \hat{u}(\tau)), \enskip \tau = t,\dots,t+T-1 \\
& p(\hat{x}(\tau), \hat{u}(\tau), \tau) \leq 0, \enskip \tau = t,\dots,t+T-1 \\
& q(\hat{x}(\tau), \hat{u}(\tau), \tau) = 0, \enskip \tau = t,\dots,t+T-1 \\
& \hat{x}(t) = x. \\
\end{aligned}\end{small}
\end{equation}
In the above formulation, the optimization variables $\hat{x}(\tau)$ and $\hat{u}(\tau)$ represent the predicted states and inputs in a future horizon $\{t,t+1,\dots,t+T\}$ with length $T\in\mathbb{N}$. The predicted trajectory is constrained by the dynamics \eqref{eq:sys} as well as some additional path constraints $p$, $q$ such as the bounds on the inputs and states or Lyapunov descent to enforce stability. Functions $\ell$ and $\ell^\mathrm{f}$ are called the stage cost and terminal cost, respectively. By solving \eqref{eq:OPC}, one executes $u(t)=\hat{u}(t)$. For simplicity it is assumed here that the states are observable; otherwise, the states can be estimated using an optimization formulation such as moving horizon estimation (MHE). For continuous-time systems, collocation techniques can be used to discretize the resulting optimal control problem into a finite-dimensional one. 
\par Now suppose that the system \eqref{eq:sys} is large-scale with its states and outputs decomposed into $n$ subsystems: $x=[x_1^\top, x_2^\top, \dots, x_n^\top]^\top$, $u=[u_1^\top, u_2^\top, \dots, u_n^\top]^\top$, and that the optimal control problem should be solved by the corresponding $n$ agents, each containing the model of its own subsystem:
\begin{equation}
x_i(\tau+1) = f_i(x_i(\tau), u_i(\tau), \{x_{ji}(\tau), u_{ji}(\tau)\}_{j\in\mathcal{P}(i)}). 
\end{equation} 
where $\{x_{ji}, u_{ji}\}$ stands for the states and inputs in subsystem $j$ (i.e., components of $x_j$ and $u_j$) that appear in the arguments of $f_i$, which comprise of the components of $f$ corresponding to the $i$-th subsystem. $\mathcal{P}_i$ is the collection of subsystems $j$ that has some inputs and outputs influencing subsystem $i$. We assume that the cost functions and the path constraints are separable:
\begin{equation}
\begin{small}\begin{aligned}
& \ell(\hat{x}, \hat{u}) = \sum_{i=1}^n \ell_i(\hat{x}_i, \hat{u}_i), \enskip \ell^\mathrm{f}(\hat{x}) = \sum_{i=1}^n \ell_i^\mathrm{f}(\hat{x}_i), \\
& p(\hat{x}, \hat{u}, \tau) = [p_1(\hat{x}_1, \hat{u}_1, \tau)^\top, \dots, p_n(\hat{x}_n, \hat{u}_n, \tau)^\top]^\top, \\
& q(\hat{x}, \hat{u}, \tau) = [q_1(\hat{x}_1, \hat{u}_1, \tau)^\top, \dots, q_n(\hat{x}_n, \hat{u}_n, \tau)^\top]^\top. \\ 
\end{aligned}\end{small}
\end{equation}

\subsection{Formulation on directed and bipartite graphs}
\par
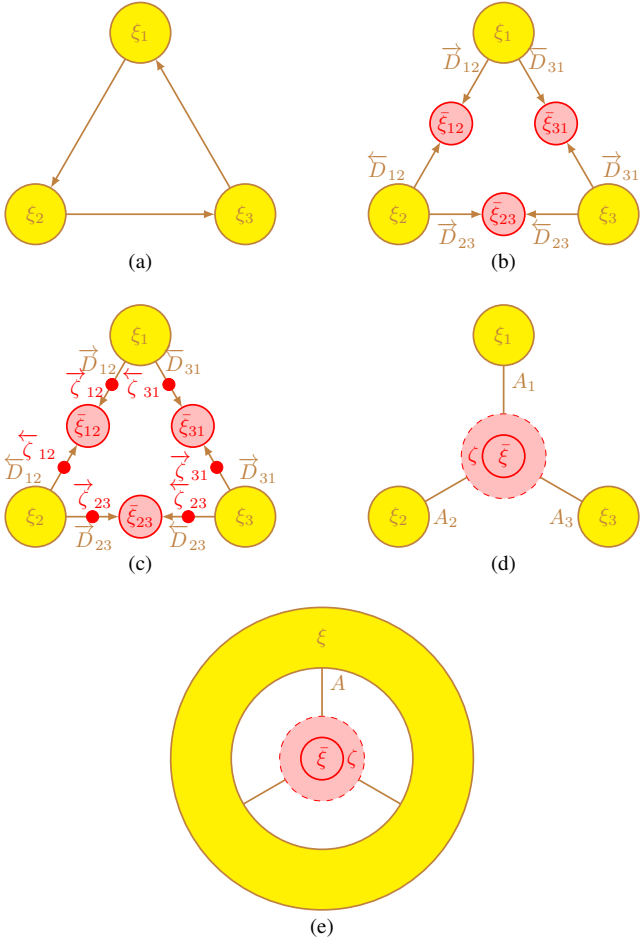
\begin{figure}
	\centering
	\resizebox{\columnwidth}{!}{
	\begin{tikzpicture}
	\draw[thick, brown, fill=yellow] (0, 2) circle [radius = 0.5] node[brown]{$\xi_1$}; 
	\draw[thick, brown, fill=yellow] (-1.7321, -1) circle [radius = 0.5] node[brown]{$\xi_2$}; 
	\draw[thick, brown, fill=yellow] (1.7321, -1) circle [radius = 0.5] node[brown]{$\xi_3$}; 
	\draw[thick, brown, -latex] (0-0.5/2, 2-0.5/2*1.7321) -- (-1.7321+0.5/2, -1+0.5/2*1.7321);
	\draw[thick, brown, -latex] (-1.7321+0.5, -1) -- (1.7321-0.5, -1);
	\draw[thick, brown, -latex] (1.7321-0.5/2, -1+0.5/2*1.7321) -- (0+0.5/2, 2-0.5/2*1.7321);
	\node at (0, -1.8) {(a)}; 
	
	\draw[thick, brown, fill=yellow] (6+0, 2) circle [radius = 0.5] node[brown]{$\xi_1$}; 
	\draw[thick, brown, fill=yellow] (6-1.7321, -1) circle [radius = 0.5] node[brown]{$\xi_2$}; 
	\draw[thick, brown, fill=yellow] (6+1.7321, -1) circle [radius = 0.5] node[brown]{$\xi_3$}; 
	\draw[thick, red, fill=pink] (6-1.7321/2, 0.5) circle [radius = 0.35] node[red]{$\bar{\xi}_{12}$}; 
	\draw[thick, red, fill=pink] (6+1.7321/2, 0.5) circle [radius = 0.35] node[red]{$\bar{\xi}_{31}$}; 
	\draw[thick, red, fill=pink] (6, -1) circle [radius = 0.35] node[red]{$\bar{\xi}_{23}$}; 
	\draw[thick, brown, -latex] (6-0.5/2, 2-0.5/2*1.7321) node[left]{$\overrightarrow{D}_{12}$} -- (6-1.7321/2+0.35/2, 0.5+0.35/2*1.7321);
	\draw[thick, brown, -latex] (6-1.7321+0.5/2, -1+0.5/2*1.7321) node[above left]{$\overleftarrow{D}_{12}$} -- (6-1.7321/2-0.35/2, 0.5-0.35/2*1.7321);
	\draw[thick, brown, -latex] (6-1.7321+0.5, -1) node[below right]{$\overrightarrow{D}_{23}$} -- (6-0.35, -1); 
	\draw[thick, brown, -latex] (6+1.7321-0.5, -1) node[below left]{$\overleftarrow{D}_{23}$} -- (6+0.35, -1);
	\draw[thick, brown, -latex] (6+1.7321-0.5/2, -1+0.5/2*1.7321) node[above right]{$\overrightarrow{D}_{31}$} -- (6+1.7321/2+0.35/2, 0.5-0.35/2*1.7321);
	\draw[thick, brown, -latex] (6+0.5/2, 2-0.5/2*1.7321) node[right]{$\overleftarrow{D}_{31}$} -- (6+1.7321/2-0.35/2, 0.5+0.35/2*1.7321) ;
	\node at (6, -1.8) {(b)}; 
	
	\draw[thick, brown, fill=yellow] (0, -5+2) circle [radius = 0.5] node[brown]{$\xi_1$}; 
	\draw[thick, brown, fill=yellow] (-1.7321, -5+-1) circle [radius = 0.5] node[brown]{$\xi_2$}; 
	\draw[thick, brown, fill=yellow] (1.7321, -5+-1) circle [radius = 0.5] node[brown]{$\xi_3$}; 
	\draw[thick, red, fill=pink] (-1.7321/2, -5+0.5) circle [radius = 0.35] node[red]{$\bar{\xi}_{12}$}; 
	\draw[thick, red, fill=pink] (1.7321/2, -5+0.5) circle [radius = 0.35] node[red]{$\bar{\xi}_{31}$}; 
	\draw[thick, red, fill=pink] (0, -5-1) circle [radius = 0.35] node[red]{$\bar{\xi}_{23}$}; 
	\draw[thick, brown, -latex] (-0.5/2, -5+2-0.5/2*1.7321) node[left]{$\overrightarrow{D}_{12}$} -- (-1.7321/2+0.35/2, -5+0.5+0.35/2*1.7321);
	\filldraw[red] (-0.5/4-1.7321/4+0.35/4, -5+2/2-0.5/4*1.7321+0.5/2+0.35/4*1.7321) circle [radius = 0.1] node[left]{$\overrightarrow{\zeta}_{12}$}; 
	\draw[thick, brown, -latex] (-1.7321+0.5/2, -5-1+0.5/2*1.7321) node[above left]{$\overleftarrow{D}_{12}$} -- (-1.7321/2-0.35/2, -5+0.5-0.35/2*1.7321);
	\filldraw[red] (-1.7321/2+0.5/4-1.7321/4-0.35/4, -5-1/2+0.5/4*1.7321+0.5/2-0.35/4*1.7321) circle [radius = 0.1] node[above left]{$\overleftarrow{\zeta}_{12}$}; 
	\draw[thick, brown, -latex] (-1.7321+0.5, -5-1) node[below right]{$\overrightarrow{D}_{23}$} -- (-0.35, -5-1); 
	\filldraw[red] (-1.7321/2+0.5/2-0.35/2, -5-1) circle [radius = 0.1] node[above]{$\overrightarrow{\zeta}_{23}$}; 
	\draw[thick, brown, -latex] (1.7321-0.5, -5-1) node[below left]{$\overleftarrow{D}_{23}$} -- (0.35, -5-1);
	\filldraw[red] (1.7321/2-0.5/2+0.35/2, -5-1) circle [radius = 0.1] node[above]{$\overleftarrow{\zeta}_{23}$}; 
	\draw[thick, brown, -latex] (1.7321-0.5/2, -5-1+0.5/2*1.7321) node[above right]{$\overrightarrow{D}_{31}$} -- (1.7321/2+0.35/2, -5+0.5-0.35/2*1.7321);
	\filldraw[red] (1.7321/2-0.5/4+1.7321/4+0.35/4, -5-1/2+0.5/4*1.7321+0.5/2-0.35/4*1.7321) circle [radius = 0.1] node[left]{$\overrightarrow{\zeta}_{31}$}; 
	\draw[thick, brown, -latex] (0.5/2, -5+2-0.5/2*1.7321) node[right]{$\overleftarrow{D}_{31}$} -- (1.7321/2-0.35/2, -5+0.5+0.35/2*1.7321);
	\filldraw[red] (0.5/4+1.7321/4-0.35/4, -5+1-0.5/4*1.7321+0.5/2+0.35/4*1.7321) circle [radius = 0.1] node[left]{$\overleftarrow{\zeta}_{31}$};
	\node at (0, -5-1.8) {(c)}; 
	
	\draw[thick, brown, fill=yellow] (6, -5+2) circle [radius = 0.5] node[brown]{$\xi_1$}; 
	\draw[thick, brown, fill=yellow] (6-1.7321, -5-1) circle [radius = 0.5] node[brown]{$\xi_2$}; 
	\draw[thick, brown, fill=yellow] (6+1.7321, -5-1) circle [radius = 0.5] node[brown]{$\xi_3$}; 
	\draw[dashed, red, fill=pink] (6, -5) circle [radius = 0.70];
	\draw[thick, red, fill=pink] (6, -5) circle [radius = 0.35] node[red]{$\bar{\xi}$}; 
	\node[red] at (5+0.5, -5) {$\zeta$};
	\draw[-, thick, brown] (6, -5+0.7) -- (6, -5+2-0.5) node[below right]{$A_1$};
	\draw[-, thick, brown] (6-0.7*1.7321/2, -5-0.7/2) -- (6-1.7321+0.5*1.7321/2, -5-1+0.5/2) node[below right]{$A_2$};
	\draw[-, thick, brown] (6+0.7*1.7321/2, -5-0.7/2) -- (6+1.7321-0.5*1.7321/2, -5-1+0.5/2) node[below left]{$A_3$};
	\node at (6, -6.8) {(d)}; 
	
	\draw[thick, brown, fill=yellow] (3, -10) circle [radius = 2.5];
	\node[brown] at (3, -10+2) {$\xi$}; 
	\draw[thick, brown, fill=white] (3, -10) circle [radius = 1.5];
	\draw[dashed, red, fill=pink] (3, -10) circle [radius = 0.70];
	\draw[thick, red, fill=pink] (3, -10) circle [radius = 0.35] node[red]{$\bar{\xi}$}; 
	\node[red] at (3+0.5, -10) {$\zeta$};
	\draw[-, thick, brown] (3, -10+0.7) -- (3, -10+2-0.5) node[below right]{$A$};
	\draw[-, thick, brown] (3-0.7*1.7321/2, -10-0.7/2) -- (3-1.7321+0.5*1.7321/2, -10-1+0.5/2);
	\draw[-, thick, brown] (3+0.7*1.7321/2, -10-0.7/2) -- (3+1.7321-0.5*1.7321/2, -10-1+0.5/2);
	\node at (3, -10-2.8) {(e)}; 
	\end{tikzpicture}
	}\caption{Graphical illustrations of the problem structure of distributed MPC.}\label{fig:structure}
\end{figure}

 To better visualize the problem structure and systematically reformulate the optimal control problem \eqref{eq:OPC} into the distributed optimization problem in the form of \eqref{eq:problem.2} for the implementation of the ELLADA algorithm, we introduce some graph-theoretic descriptions of optimization problems \cite{daoutidis2019decomposition}. For problem \eqref{eq:OPC}, we first define a directed graph (digraph), which is a straightforward characterization of the relation of mutual impact among the subsystem models.
\begin{definition}[Digraph]\label{def:digraph}
	The digraph of system \eqref{eq:sys} under the decomposition $x=[x_1^\top, x_2^\top, \dots, x_n^\top]^\top$ and $u=[u_1^\top, u_2^\top, \dots, u_n^\top]^\top$ is $\mathcal{G}_1 = \{\mathcal{V}_1, \mathcal{E}_1\}$ with nodes $\mathcal{V}_1 = \{1,2,\dots,n\}$ and edges $\mathcal{E}_1 = \{(j,i) | j \in \mathcal{P}(i)\}$. If $(i,j)\in\mathcal{E}_1$, i.e., $j\in\mathcal{P}(i)$, we say that $j$ is a {parent} of $i$ and $i$ is a {child} of $j$ (denoted as $i\in\mathcal{C}(j)$). 
\end{definition}
Then under the decomposition, \eqref{eq:OPC} can be written as
\begin{equation}
\begin{small}\begin{aligned}
\min \enskip & \sum_{i\in\mathcal{V}_1} J_i = \sum_{i\in\mathcal{V}} \sum_{\tau=t}^{t+T-1} \ell_i(\hat{x}_i(\tau), \hat{u}_i(\tau)) + \ell_i^\mathrm{f}(\hat{x}_i(t+T)) \\
\mathrm{s.t.} \enskip & \hat{x}_i(\tau+1) = f_i(\hat{x}_i(\tau), \hat{u}_i(\tau), \{\hat{x}_{ji}(\tau), \hat{u}_{ji}(\tau)\}_{j\in\mathcal{P}(i)}), \\
& p_i(\hat{x}_i(\tau), \hat{u}_i(\tau), \tau) \leq 0, \enskip \tau = t,\dots,t+T-1, \enskip i\in\mathcal{V}_1 \\
& q_i(\hat{x}_i(\tau), \hat{u}_i(\tau), \tau) = 0, \enskip \tau = t,\dots,t+T-1, \enskip i\in\mathcal{V}_1 \\
& \hat{x}_i(t) = x_i,  \enskip i\in\mathcal{V}_1, \\
\end{aligned}\end{small}
\end{equation}
We denote the variables of the $i$-th agent as 
\begin{equation}
\begin{small}\begin{aligned}
\xi_i = [& \hat{x}_i(t)^\top, \hat{u}_i(t)^\top, \dots, \hat{x}_i(t+T-1)^\top, \hat{u}_i(t+T-1)^\top, \\
& \hat{x}_i(t+T)^\top, \{\hat{x}_{ji}(t)^\top, \hat{u}_{ji}(t)^\top\}_{j\in\mathcal{P}(i)}, \\
& \dots, \{\hat{x}_{ji}(t+T-1)^\top, \hat{u}_{ji}(t+T-1)^\top\}_{j\in\mathcal{P}(i)}]^\top,
\end{aligned}\end{small}
\end{equation}
in which the variables related to the $j$-th subsystem are denoted as $\xi_{ji}$. Since $\xi_{ji}$ is a part of the predicted states and inputs from subsystem $j$, i.e., some components of $\xi_j$, the interactions between the parent $j$ and the child $i$ be captured by a matrix $\overrightarrow{D}_{ji}$ with exactly one unit entry (``1") on every row: $\xi_{ji} = \overrightarrow{D}_{ji}\xi_j$, where the right arrow represents the impact of the parent subsystem $j$ on the child subsystem $i$. By denoting the model and path constraints in agent $i$ as $\xi_i\in\Xi_i$, the optimal control problem \eqref{eq:sys} is expressed in a compact way as follows:
\begin{equation}\label{eq:OPCa}
\begin{small}\begin{aligned}
\min\enskip & \sum_{i\in\mathcal{V}_1} J_i(\xi_i) \\
\mathrm{s.t.}\enskip & \xi_i\in\Xi_i, \enskip i\in\mathcal{V}_1, \enskip \xi_{ji} = \overrightarrow{D}_{ji}\xi_j, \enskip (j,i)\in\mathcal{E}_1. \\
\end{aligned}\end{small}
\end{equation}
This is an optimization problem defined on a \emph{directed graph}. An illustration for a simple case when $\mathcal{E}_1=\{(1,2), (2,3), (3,1)\}$ is shown in Fig. \ref{fig:structure}(a). 

\par Although it is natural to represent the interactions among the subsystems in a digraph, performing distributed optimization on digraphs where the agents communicate among themselves without a coordinator can be challenging. For example, it is known that the ADMM algorithm, which behaves well for distributed optimization with 2 blocks of variables, can become divergent when directly extended to multi-block problems \cite{chen2016direct}. Hence we construct such a 2-block architecture by using a \emph{bipartite graph}. 
\begin{definition}[Bipartite graph]\label{defn:bipartite}
	The bipartite graph of system \eqref{eq:sys} $\mathcal{G}_2$ is constructed from the digraph $\mathcal{G}_1$ by taking both the  nodes and edges as the new nodes, and adding an edge between $i\in\mathcal{V}_1$ and $e\in\mathcal{E}_1$ if $i$ is the head or tail of $e$ in the digraph, i.e., $\mathcal{G}_2=(\mathcal{V}_2, \mathcal{E}_2)$ with $\mathcal{V}_2=\mathcal{V}_1 \cup \mathcal{E}_1$, $\mathcal{E}_2=\{(i,e)|i\in\mathcal{V}_1, e\in\mathcal{E}_1, e=(i,j), j\in\mathcal{C}(i) \text{ or } e=(j,i), j\in\mathcal{P}(i)\}$.
\end{definition} 
Such a graph is bipartite since any edge is between a node of $\mathcal{V}_1$ and a node of $\mathcal{E}_1$. 
\par We note that the last line of \eqref{eq:OPCa} corresponds to the digraph edges $\mathcal{E}_1$. In the bipartite graph, these edges should become nodes and hence new groups of variables should be associated with them. For this purpose, we simply need to pull out $\xi_{ji}$ as overlapping variables $\bar{\xi}_{ji}$, and add the constraint that $\bar{\xi}_{ji}$ are some selected components of $\xi_i$: $\xi_{ji}=\overleftarrow{D}_{ji}\xi_i$:
\begin{equation}\label{eq:OPCb}
\begin{small}\begin{aligned}
\min\enskip & \sum_{i\in\mathcal{V}_1} J_i(\xi_i) \\
\mathrm{s.t.}\enskip & \xi_i\in\Xi_i, \enskip i\in\mathcal{V}_1, \enskip
\bar{\xi}_{ji} = \overrightarrow{D}_{ji}\xi_j = \overleftarrow{D}_{ji}\xi_j, \enskip (j,i)\in\mathcal{E}_1 \\
\end{aligned}\end{small}
\end{equation}
In \eqref{eq:OPCb}, variables $\xi_i$ ($i\in\mathcal{V}_1$) and $\bar{\xi}_{ji}$ ($(j,i)\in\mathcal{E}_1$) are defined on the nodes of the bipartite graph, and the constraints captured by the matrices $\overrightarrow{D}_{ji}$ and $\overleftarrow{D}_{ji}$ correspond to the bipartite edges $(j, (j,i))$ and $(i, (j,i))$, respectively. We may also write the last line of \eqref{eq:OPCb} as
\begin{equation}
\bar{\xi}_e = D_{ie}\xi_i, \enskip (i,e)\in\mathcal{E}_2.
\end{equation}
Therefore \eqref{eq:OPCb} is an optimization problem on the bipartite graph. An illustration is given in Fig. \ref{fig:structure}(b). Under this reformulation, the problem structure becomes a 2-block one --  distributed agents $i=1,\dots,N$ manage the decision variables $\xi_i$, $\mathcal{V}_1$ in parallel without interference, and the coordinator regulates the agents by using overlapping variables $\bar{\xi}_e$, $e\in\mathcal{E}_1$.

\subsection{Reformulation with slack variables}
\par It is known that a key condition for distributed optimization in the context of the ADMM algorithm to converge is that one block of variables can always be made feasible given the other block \cite{wang2019global}. Unfortunately this condition is not always met by the problem \eqref{eq:OPCb}. For example, given $\xi_1$ and $\xi_2$, there may not be a $\bar{\xi}_{12}$ satisfying both $\bar{\xi}_{12} = \overrightarrow{D}_{12}\xi_1$ and $\bar{\xi}_{12} = \overleftarrow{D}_{12}\xi_2$. To deal with this issue, it was proposed to associate with each linear constraint in \eqref{eq:OPCb}, namely each edge in the bipartite graph, a slack variable $\zeta_{ie}$ (e.g., \cite{sun2019two}):
\begin{equation}\label{eq:OPCc}
\begin{small}\begin{aligned}
\min\enskip & \sum_{i\in\mathcal{V}_1} J_i(\xi_i) \\
\mathrm{s.t.}\enskip & \xi_i\in\Xi_i, \enskip i\in\mathcal{V}_1 \\
& D_{ie}\xi_i - \bar{\xi}_e + \zeta_{ie} = 0, \enskip (i,e)\in\mathcal{E}_2 \\
& \zeta_{ie} = 0, \enskip (i,e)\in\mathcal{E}_2. \\
\end{aligned}\end{small}
\end{equation}
Similar to the notation for $D$, we write $\zeta_{ie}$ as $\overrightarrow{\zeta}_{ij}$ if $e=(i,j)$ and $\overleftarrow{\zeta}_{ij}$ if $e=(j,i)$. Such a problem structure is graphically illustrated in Fig. \ref{fig:structure}(c). 
\par Finally, we stack all the subscripted variables into $\xi$, $\bar{\xi}$, $\zeta$ in a proper ordering of $i\in\mathcal{V}_1$, $e\in\mathcal{E}_1$, and $(i,e)\in\mathcal{E}_2$. The matrices $D_{ie}$ are stacked in a block diagonal pattern in the same ordering of $(i,e)\in\mathcal{E}_2$ into $A$. The appearance of $\bar{\xi}_e$ in the equality constraints is represented by a matrix $B$ (satisfying $B^\top B = 2I$). We write the objective function as $J(\xi)$, and the set constraints $\Xi_i$ are lumped into a Cartesian product $\Xi=\times_{i\in\mathcal{V}_1} \Xi_i$. Finally, we reach a compact formulation for \eqref{eq:OPCc}:
\begin{equation}
\begin{small}\begin{aligned}
\min \enskip & J(\xi) \\
\mathrm{s.t.} \enskip & \xi \in \Xi, \enskip A\xi + B\bar{\xi} + \zeta = 0, \enskip \zeta = 0 \\
\end{aligned}\end{small}
\end{equation}
Such an architecture is shown in Figs. \ref{fig:structure}(d) and \ref{fig:structure}(e). The variables $\bar{\xi}$ and $\zeta$ belong to the coordinator (marked in red), and $\xi$ is in the distributed agents.

\subsection{Implementation of ELLADA}\label{subsec:A}
\par Clearly, the optimal control problem formulated as \eqref{eq:OPCc} is a special form of \eqref{eq:problem.2} with $\xi$, $\bar{\xi}$ and $\zeta$ rewritten as $x$, $\bar{x}$ and $z$, respectively, and $g(\bar{x})=0$, $\bar{\mathcal{X}}$ equal to the entire Euclidean space. As long as the cost function $J$ is lower bounded (e.g., a quadratic cost), Algorithm \ref{alg:A3} is applicable to \eqref{eq:OPCc}, where the operations on $\bar{x}$, $z$, $y$ are performed by the coordinator, and the operations on $x$ is handled by the distributed agents. Specifically, 
\begin{itemize}
    \item The update steps of $\bar{x}, z, y$ (Lines 10--13, 15, 16) and the entire Anderson acceleration (Lines 17--26) belong to the coordinator. The updates of penalty parameters and outer-layer dual variables $\lambda$ (Lines 31) should also be performed by the coordinator. The conditions for $\epsilon_1^{k}, \epsilon_2^k, \epsilon_3^k$ and $\epsilon_1, \epsilon_2, \epsilon_3$ are checked by the coordinator.
    \item The distributed agents are responsible for carrying out a trial $x$-update step for the Anderson acceleration (Line 9) as well as the plain $x$-update (Line 14). The conditions and updates for $\epsilon_4^{k,r}, \epsilon_5^{k,r}$, $\epsilon_4^k, \epsilon_5^k$, and $\epsilon_4, \epsilon_5, \epsilon_6$ are checked by the agents. 
\end{itemize}

\newenvironment{rcases}
  {\left.\begin{small}\begin{aligned}}
  {\end{aligned}\end{small}\right\rbrace}
\newcommand{\parallelsum}{\mathbin{\!/\mkern-5mu/\!}}

\par When executing the updates, the agents need the values of $B\bar{x}+z+y/\rho$ to add to $Ax$, and the coordinator needs the value of $Ax$ from the agents. When the variables $x$ are distributed into agents $x_1,\dots,x_n$, and the equality constraints between the agents and the coordinator is expressed on a bipartite graph:
\begin{equation}
    D_{ie}x_i - \bar{x}_e + z_{ie} = 0, \enskip (i,e)\in\mathcal{E}_2,
\end{equation}
the communication of $Ax$ and $B\bar{x}+z+y/\rho$ takes place in a distributed and parallel way, i.e., 
the $i$-th agent obtains the information of $-\bar{x}_e + z_{ie} + y_{ie}/\rho$ for all $e$ such that $(i,e)\in\mathcal{E}_2$ from the coordinator. The coordinator, based on inter-subsystem edges $e$ in the digraph, obtains the information of $D_{ie}x_i$ for all related agents $i$. When the objective function and $\mathcal{X}$ are separable $f(x) = \sum_{i=1}^n f_i(x_i)$, $\mathcal{X} = \mathcal{X}_1 \times \dots \times \mathcal{X}_n$, based on such distributed and parallel communication, the optimization problem
\begin{equation}
\begin{small}\begin{aligned}
    \min \enskip & f(x) - b\sum_{c=1}^{C_\phi}\ln(-\phi_c(x)) + \frac{\rho}{2}\left\| Ax + B\bar{x}+ z + \frac{y}{\rho} \right\|^2 \\
    \mathrm{s.t.} \enskip & \psi(x) = 0 \\
\end{aligned}\end{small}
\end{equation}
in an $x$-update step can be solved in a distributed and parallel manner:
\begin{equation}\label{eq:x.update.distributed}
\begin{rcases}
    \min_{x_i} \enskip & f_i(x_i) - b\sum_{c=1}^{C_{\phi, i}}\ln(-\phi_{c, i}(x_i)) \\
    & +\frac{\rho}{2}\sum_{\{e|(i,e)\in\mathcal{E}_2\}}\left\| D_{ie}x_i - \bar{x}_e + z_{ie} + \frac{y_{ie}}{\rho} \right\|^2 \\
    \mathrm{s.t.} \enskip & \psi_i(x_i) = 0 \\
\end{rcases}
\text{ $\parallelsum$ for $i$.}
\end{equation}
Similarly, the $\bar{x}$-update with the $G$-mapping is in parallel for its components $e$, if $\bar{\mathcal{X}}$ is separable, i.e., if $\bar{\mathcal{X}}$ is a closed hypercube (whether bounded or unbounded), and if $g$ is also separable. That is, $\bar{x}$-update can be expressed as
\begin{equation}
\begin{rcases}
    \min_{\bar{x}_i} \enskip & g_i(\bar{x}_i) + \frac{\rho}{2}\sum_{\{i|(i,e)\in\mathcal{E}_2\}}\left\| D_{ie}x_i - \bar{x}_e + z_{ie} + \frac{y_{ie}}{\rho} \right\|^2  \\
    \mathrm{s.t.} \enskip & \bar{x}_i \in \bar{\mathcal{X}}_i \\
\end{rcases}
\text{ $\parallelsum$ for $e$.}
\end{equation}
The $z$ and $y$ updates are in parallel for the edges $(i,e)$ on the bipartite graph. 

\par In Algorithm \ref{alg:A3}, the procedures are written such that in each iteration, the update steps are carried out in sequence. This requires a synchronization of all the agents $i$ and the coordinating elements $e$ and $(i,e)$. For example, for the $x$-update, every distributed agent needs to create a ``finish" signal after solving $x_i$ in \eqref{eq:x.update.distributed} and send it to the coordinator. Only after the coordinator receives the ``finish" signals from all the distributed agents can the $\bar{x}$-update be carried out. Due to the possible computational imbalance among the agents and the coordinator, such synchronization implies that faster updates must idle for some time to wait for slower ones. In fact, the convergence properties of the ELLADA algorithm do not rely on the synchronization. Even when the inner iterations are asynchronous, the update steps still contribute to the convergence of the barrier augmented Lagrangian and hence result in convergence to KKT conditions. The only exception is that under Anderson acceleration, the steps for generating the candidate of accelerated updates are allocated to another coordinator and another set of distributed agents, and they should communicate to make the decision on executing the accelerations.

\section{Application to a Quadruple Tank Process}\label{sec:tanks}
\par The quadruple tank process is a simple benchmark process for distributed model predictive control \cite{johansson2000quadruple} with 4 states (water heights in the 4 tanks) and 2 inputs (flow rates from the reservoir). The dynamic model is written as follows:
\begin{equation}
\begin{small}
\begin{aligned}
    \dot{h}_1 &= -\frac{a_1}{A_1}\sqrt{h_1} + \frac{a_3}{A_1}\sqrt{h_3} + \frac{\gamma_1k_1}{A_1}v_1 \\
    \dot{h}_2 &= -\frac{a_2}{A_2}\sqrt{h_2} + \frac{a_4}{A_2}\sqrt{h_4} + \frac{\gamma_2k_2}{A_2}v_2 \\
    \dot{h}_3 &= -\frac{a_3}{A_3}\sqrt{h_3} + \frac{(1-\gamma_2)k_2}{A_3}v_2 \\
    \dot{h}_4 &= -\frac{a_4}{A_4}\sqrt{h_4} + \frac{(1-\gamma_1)k_1}{A_4}v_1.
\end{aligned}
\end{small}
\end{equation}
\begin{table}[!t]
    \centering
    \caption{Parameters and nominal steady state}
    \label{tab:parameters}
    \begin{tabular}{|cc|cc|}
        \hline
        Parameter & Value & Parameter & Value \\
        $A_1$, $A_3$ & 28 & $a_1$, $a_3$ & 3.145 \\
        $A_2$, $A_4$ & 32 & $a_2$, $a_4$ & 2.525 \\
        $\gamma_1$ & 0.43 & $k_1$ & 3.14 \\
        $\gamma_2$ & 0.34 & $k_2$ & 3.29 \\
        \hline 
        Input & Value & Input & Value \\
        $v_1$ & 3.15 & $v_2$ & 3.15 \\
        \hline
        State & Value & State & Value \\
        $h_1$ & 12.44 & $h_2$ & 13.17 \\
        $h_3$ & 4.73 & $h_4$ & 4.99 \\
        \hline
    \end{tabular}
\end{table}
Other parameter values and the nominal steady state are given in Table \ref{tab:parameters}. The process is considered to have 2 subsystems, one containing tanks 1 and 4 and the other containing tanks 2 and 3. Each subsystem has 2 states, 1 input and 1 upstream state. We first design a centralized MPC with quadratic objective function for each tank, and bounds on the inputs $2.5 \leq v_1, v_2 \leq 3.5$. We first decide through the simulation of centralized MPC that a receding horizon of $T=400$ with sampling time $\delta t = 10$ is appropriate. (The computations are performed using the Python module \texttt{pyomo.dae} with an IPOPT solver \cite{nicholson2018pyomo}.) 

\begin{figure*}[!t]
    \centering
    \includegraphics[width = \textwidth]{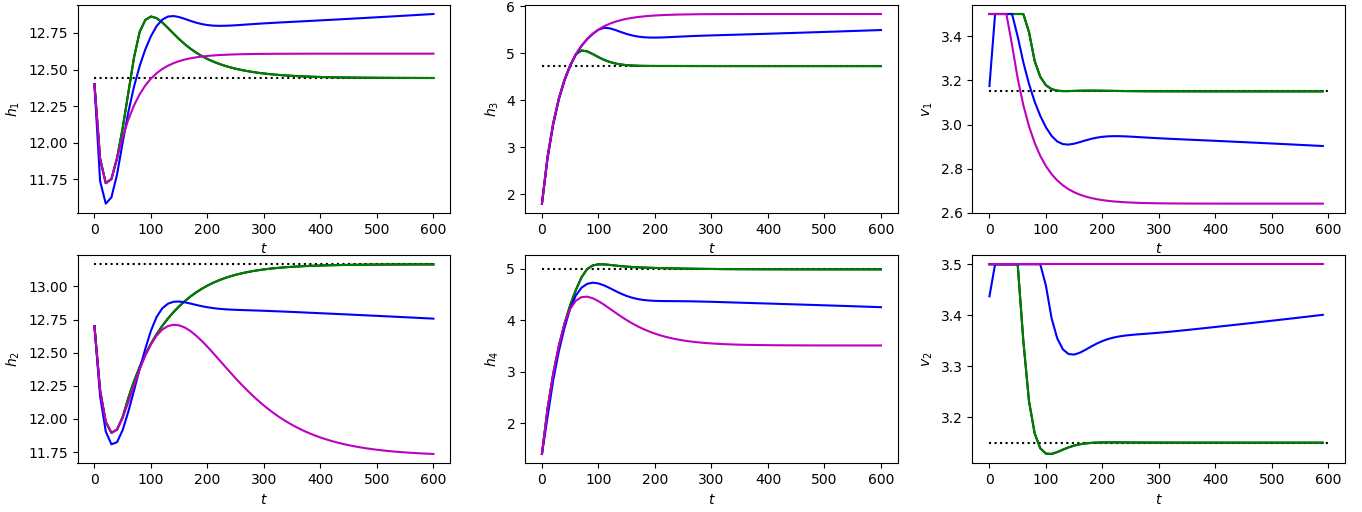}
    \caption{Closed-loop trajectories under traditional MPC controllers.}
    \label{fig:quadruple.traditional}
\end{figure*}
\par The closed-loop trajectories under the traditional MPC controllers, including a centralized MPC (black), a semi-centralized MPC where the inputs are iteratively updated based on predictions over the entire process (green), a decentralized MPC (blue), and a distributed MPC with only state feedforwarding among the agents (purple), are shown in Fig. \ref{fig:quadruple.traditional}. It was observed that a semi-centralized MPC based on system-wide prediction maintains the control performance, yielding trajectories overlapping with those of the centralized MPC. However, the state-feedforward distributed MPC without sufficient coordination accounting for the state interactions results in unsatisfactory control performance, whose ultimate deviation from the steady state is even larger than the decentralized MPC without any communication between the controllers. 

\begin{figure*}
    \centering
    \includegraphics[width = \textwidth]{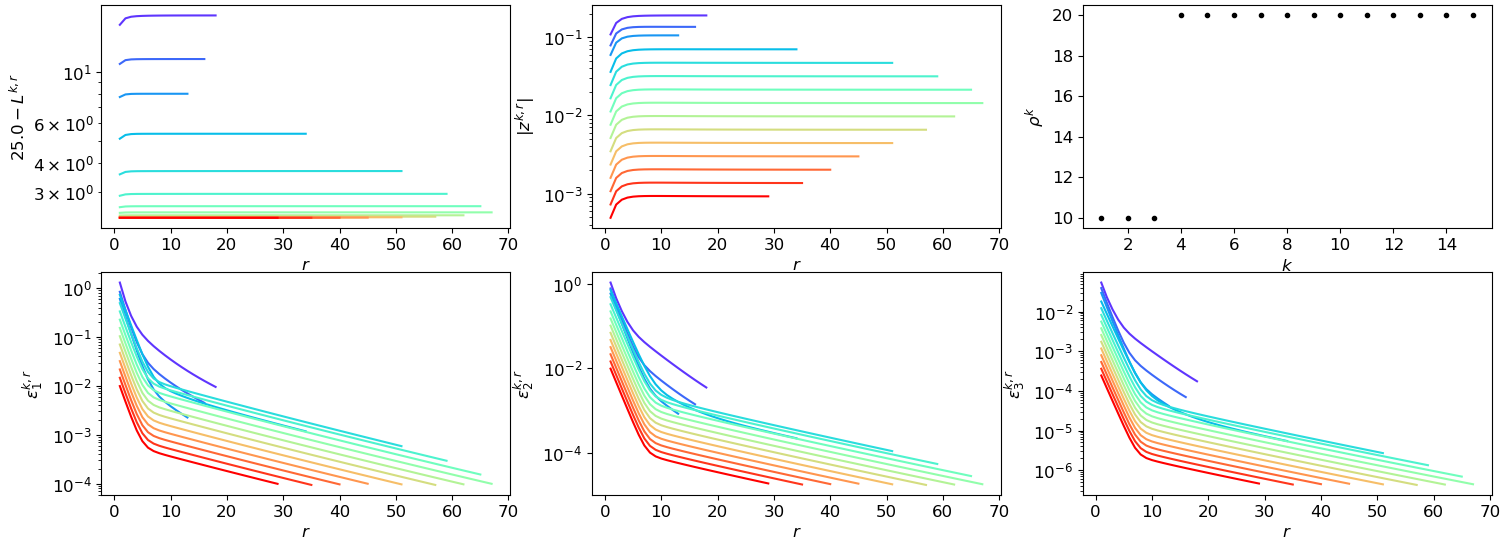}
    \caption{Solution results of the ELL algorithm.}\label{fig:ELL}
\end{figure*}
\par Next we use the proposed ELLADA algorithm for distributed nonlinear MPC of the process. We first examine the basic ELL algorithm (Algorithm \ref{alg:A1}) by solving the corresponding distributed MPC problem at a state with $h_1 = 12.6$, $h_2 = 12.4$, $h_3 = 5.0$, $h_4 = 4.5$, where we set $\omega = 0.75$, $\gamma = 2$, $\epsilon_1^k = \epsilon_2^k = 10^{-2}/2^{k-1}$, $\epsilon_3^k = 10^{-1}/2^{k-1}$, $\epsilon_1 = \epsilon_2 = 10^{-4}$, $\epsilon_3 = 10^{-3}$ and $\overline{\lambda} = -\underline{\lambda} = 10$ (in an element-wise sense) through empirical tuning. The solution results in terms of the variation of the augmented Lagrangian $L^{k,r}$, the violations to the KKT conditions  $\epsilon_{1,2,3}^{k,r}$, and penalty parameters $\rho^k$ throughout the inner and outer iterations are presented in Fig. \ref{fig:ELL}, where the rainbow colormap from blue to red colors stand for increasing outer iteration number. In accordance to the conclusion of Lemma \ref{lemma:descent}, the augmented Lagrangian is monotonically decreasing in each outer iterations and remains upper bounded, which guarantees the convergence of the algorithm. Using the ELL algorithm for the afore-mentioned closed-loop MPC simulation, the resulting trajectories are found identical to those of the centralized control, which corroborates the theoretical property of the algorithm of converging to the set of stationary solutions.

\par
\begin{figure}
    \centering
    \includegraphics[width = \columnwidth]{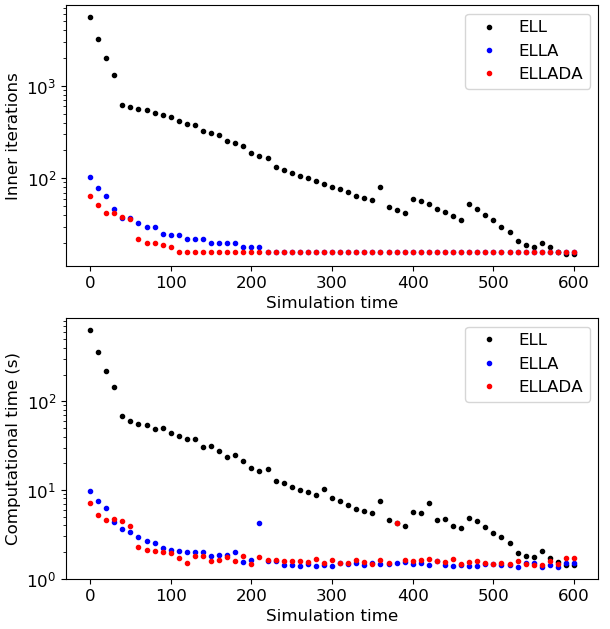}
    \caption{Iteration and computational time under ELL, ELLA and ELLADA algorithms.}\label{fig:ELLADA}
\end{figure}

 With the preserved control performance of the ELL algorithm, we seek to improve its computational efficiency with the ELLA and ELLADA algorithms (Algorithms \ref{alg:A2} and \ref{alg:A3}). In ELLA, the tolerances for approximate NLP solution are set as $\epsilon_1 = \epsilon_2 = \epsilon_4 = 10^3\epsilon_3 = 1$, $\epsilon_1^k = \epsilon_2^k = 10^3\epsilon_3^k = \epsilon_4^k = 100/2^{k-1}$, $\epsilon_4^{k,r} = 10^3 \epsilon_5^{k,r} = \max(\epsilon_4^k, 40(\epsilon_1^{k,r})^2)$. The barrier constants are updated throughout outer iterations according to $\|z\|$ according to $b^{k+1} = \min(10^{-1}, \max(10^{-4}, 25(\epsilon_3^k)^2))$. Compared to ELL, the accumulated number of iterations and computational time of ELLA are reduced by over an order of magnitude. To seek for better computational performance, we apply the ELLADA algorithm, where we set $M = 10$, $\sigma = 1$, $\eta_L = \eta_{\tilde{w}} = 0.01$, $\eta_\theta = 0.5$, $\eta_w = 0.05$. This further reduces the number of iterations and computational time. These results are shown in Fig. \ref{fig:ELLADA}. 

\par Compared to the basic ELL algorithm, ELLADA achieves acceleration by approximately 18 times in terms of iterations and 19 times in computational time for the entire simulation time span. These improvements are more significant when the states are far from the target steady state (43 and 45 times, respectively, for the first 1/6 of the simulation). We note that the improvement from ELLA to ELLADA by using the Anderson scheme is not an order-of-magnitude one mainly because each outer iteration needs only a few number of inner iterations, leaving little space for further acceleration (e.g., for the first sampling time, 12 outer iterations including only 102 inner iterations are needed in ELLA, and in ELLADA, 61 inner iterations are needed). Under the accelerations, ELLADA returns the identical solution to the centralized optimization, thus preserving the control performance of the centralized MPC.

\section{Conclusions and Discussions}\label{sec:conclusion}
\par We have proposed a new algorithm for distributed optimization allowing nonconvex constraints, which simultaneously guarantees convergence under mild assumptions and achieves fast computation. Specifically, convergence is established by adopting a two-layer architecture. In the outer layer, the slack variables are tightened using the method of multipliers, and the inequalities are handled using a barrier technique. In the inner layer, ADMM iterations are performed in a distributed and coordinated manner. Approximate NLP solution and Anderson acceleration techniques are integrated into inner iterations for computational acceleration. 
\par Such an algorithm is generically suitable for distributed nonlinear MPC. The advantages include:
\begin{itemize}
    \item Arbitrary input and state couplings among subsystems are allowed. No specific pattern is required a priori. 
    \item The convergence property of the algorithm towards a stationary point is theoretically guaranteed, and its performance can be monitored throughout iterations. 
    \item Equality-constrained NLP solvers can be used only as a subroutine. No internal modification of solvers is needed, and the choice of any appropriate solver is flexible.
    \item Asynchronous updates are allowed without affecting the convergence properties. 
    \item Although motivated with a nominal optimal control problem, the algorithm could be suitable for more intricate MPC formulations such as stochastic/robust MPC or sensitivity-based advance-step MPC.
\end{itemize}
\par The application of the ELLADA algorithm on the distributed nonlinear MPC of a quadruple tank process has already shown its improved computational performance compared to the basic convergent Algorithms \ref{alg:A1} and \ref{alg:A2}, and improved control performance compared to the decentralized MPC and distributed MPC without accounting for state interactions. Of course, due to the small size of the specific benchmark process, the control can be realized easily with a centralized MPC. A truly large-scale control problem is more suitable to demonstrate the effectiveness of our algorithm, and this shall be presented in an upcoming separate paper.

\appendices
\section{Proof of Lemma \ref{lemma:descent}}\label{app:A}
\par First, since $x^{k,r+1}$ is chosen as the minimizer of the augmented Lagrangian with respect to $x$ (Line 9, Algorithm \ref{alg:A1}), the update of $x$ leads to a decrease in $L$:
\begin{equation}\label{eq:A1.1}
\begin{small}\begin{aligned}
    L(x^{k,r+1}, \bar{x}^{k,r}, z^{k,r}, y^{k,r}) \leq L(x^{k,r}, \bar{x}^{k,r}, z^{k,r}, y^{k,r}).
\end{aligned}\end{small}
\end{equation}

\par Second, we consider the decrease resulted from $\bar{x}$-update:
\begin{equation}\label{eq:A1.2.1}
    \begin{small}\begin{aligned}
    & L(x^{k,r+1}, \bar{x}^{k,r+1}, z^{k,r}, y^{k,r}) - L(x^{k,r+1}, \bar{x}^{k,r}, z^{k,r}, y^{k,r}) \\
    & = g(\bar{x}^{k,r+1}) - g(\bar{x}^{k,r}) + y^{k,r\top}(B\bar{x}^{k,r+1}-B\bar{x}^{k,r}) \\
    & \quad + \frac{\rho^k}{2} \|Ax^{k,r+1} + B\bar{x}^{k,r+1} + z^{k,r}\|^2 \\
    & \quad - \frac{\rho^k}{2} \|Ax^{k,r+1} + B\bar{x}^{k,r} + z^{k,r}\|^2  \\
    & = g(\bar{x}^{k,r+1}) - g(\bar{x}^{k,r}) - \frac{\rho^k}{2}\|B\bar{x}^{k,r+1}-B\bar{x}^{k,r}\|^2 \\
    & \quad -\rho^k (\bar{x}^{k,r} - \bar{x}^{k,r+1})^\top B^\top \left( Ax^{k,r+1} + B\bar{x}^{k,r+1} + z^{k,r} + \frac{y^{k,r}}{\rho^k} \right).
    \end{aligned}\end{small}
\end{equation}
The minimization of $\bar{x}$ (Line 10, Algorithm \ref{alg:A1}) should satisfy the optimality condition
\begin{equation}
    \begin{small}\begin{aligned}
    0 \in & \rho^k B^\top \left( Ax^{k,r+1} + B\bar{x}^{k,r+1} + z^{k,r} + \frac{y^{k,r}}{\rho^k} \right) \\
    & + \partial g(\bar{x}^{k,r+1}) + \mathcal{N}_{\bar{\mathcal{X}}}(\bar{x}^{k,r+1}),
    \end{aligned}\end{small}
\end{equation}
i.e., there exist vectors $v_1\in\partial g(\bar{x}^{k,r+1})$ and $v_2 \in \mathcal{N}_{\bar{\mathcal{X}}}(\bar{x}^{k,r+1})$ with 
\begin{equation}
\begin{small}\begin{aligned}
    \rho^k B^\top \left( Ax^{k,r+1} + B\bar{x}^{k,r+1} + z^{k,r} + \frac{y^{k,r}}{\rho^k} \right) = -v_1 - v_2.
\end{aligned}\end{small}
\end{equation}
Since $v_1\in\partial g(\bar{x}^{k,r+1})$ and $g$ is convex, $v_1^\top(\bar{x}^{k,r} - \bar{x}^{k,r+1}) \leq g(\bar{x}^{k,r}) - g(\bar{x}^{k,r+1})$. And $v_2 \in \mathcal{N}_{\bar{\mathcal{X}}}(\bar{x}^{k,r+1})$ implies $v_2^\top(\bar{x}^{k,r} - \bar{x}^{k,r+1})\leq 0$. Hence
\begin{equation}\label{eq:A1.2.2}
\begin{small}\begin{aligned}
    &\rho^k (\bar{x}^{k,r} - \bar{x}^{k,r+1})^\top B^\top \left( Ax^{k,r+1} + B\bar{x}^{k,r+1} + z^{k,r} + \frac{y^{k,r}}{\rho^k} \right) \\
    & = -v_1^\top(\bar{x}^{k,r} - \bar{x}^{k,r+1}) - v_2^\top (\bar{x}^{k,r} - \bar{x}^{k,r+1}) \\
    & \geq -(g(\bar{x}^{k,r}) - g(\bar{x}^{k,r+1})).
\end{aligned}\end{small}
\end{equation}
Substituting the above inequality in \eqref{eq:A1.2.1}, we obtain
\begin{equation}\label{eq:A1.2}
\begin{small}\begin{aligned}
    L(x^{k,r+1}, \bar{x}^{k,r+1}, z^{k,r}, y^{k,r}) \leq L(x^{k,r+1}, \bar{x}^{k,r}, z^{k,r}, y^{k,r}) \\
    - \frac{\rho^k}{2}\|B\bar{x}^{k,r+1}-B\bar{x}^{k,r}\|^2.
\end{aligned}\end{small}
\end{equation}

\par Third, we consider the decrease resulted from $z$- and $y$-updates:
\begin{equation}\label{eq:A1.3.1}
    \begin{small}\begin{aligned}
    & L(x^{k,r+1}, \bar{x}^{k,r+1}, z^{k,r+1}, y^{k,r+1}) - L(x^{k,r+1}, \bar{x}^{k,r+1}, z^{k,r}, y^{k,r}) \\
    & = \lambda^{k\top}(z^{k,r+1}-z^{k,r}) + \frac{\beta^k}{2}(\|z^{k,r+1}\|^2 - \|z^{k,r}\|^2) \\
    & \quad + y^{k,r+1\top}(Ax^{k,r+1} + B\bar{x}^{k,r+1} + z^{k,r+1}) \\
    & \quad - y^{k,r\top}(Ax^{k,r+1} + B\bar{x}^{k,r+1} + z^{k,r}) \\
    & \quad + \frac{\rho^k}{2}\|Ax^{k,r+1} + B\bar{x}^{k,r+1} + z^{k,r+1}\|^2 \\
    & \quad - \frac{\rho^k}{2}\|Ax^{k,r+1} + B\bar{x}^{k,r+1} + z^{k,r}\|^2.
    \end{aligned}\end{small}
\end{equation}
Since $\upsilon(z; \lambda, \beta)=\lambda^\top z + \frac{\beta}{2}\|z\|^2$ is a convex function, whose gradient is $\nabla\upsilon(z; \lambda, \beta) = \lambda + \beta z$, 
\begin{equation}
    \begin{small}\begin{aligned}
    \upsilon(z^{k,r+1}; \lambda^k, \beta^k) - \upsilon(z^{k,r}; \lambda^k, \beta^k) \\
    \leq (\lambda^k + \beta^k z^{k,r+1})^\top(z^{k,r+1}-z^{k,r}),
    \end{aligned}\end{small}
\end{equation}
From Line 11 of Algorithm \ref{alg:A1} it can be obtained
\begin{equation}\label{eq:A1.3.2}
\begin{small}\begin{aligned}
    \lambda^k + \beta z^{k,r+1} = -y^{k,r+1}. 
\end{aligned}\end{small}
\end{equation}
Substituting into \eqref{eq:A1.3.1}, we obtain
\vspace{-12pt}
\begin{center}
\begin{small}
\begin{align}\label{eq:A1.3.3}
    & L(x^{k,r+1}, \bar{x}^{k,r+1}, z^{k,r+1}, y^{k,r+1}) - L(x^{k,r+1}, \bar{x}^{k,r+1}, z^{k,r}, y^{k,r}) \nonumber \\
    & \leq (y^{k,r+1}- y^{k,r})^\top(Ax^{k,r+1} + B\bar{x}^{k,r+1} + z^{k,r}) \nonumber \\
    & \quad + \frac{\rho^k}{2}\|Ax^{k,r+1} + B\bar{x}^{k,r+1} + z^{k,r+1}\|^2 \nonumber \\
    & \quad - \frac{\rho^k}{2}\|Ax^{k,r+1} + B\bar{x}^{k,r+1} + z^{k,r}\|^2 \\
    & = \frac{\rho^k}{2}(Ax^{k,r+1} + B\bar{x}^{k,r+1} + z^{k,r+1})^\top(Ax^{k,r+1} + B\bar{x}^{k,r+1} + z^{k,r}) \nonumber \\
    & \quad + \frac{\rho^k}{2}\|Ax^{k,r+1} + B\bar{x}^{k,r+1} + z^{k,r+1}\|^2 \nonumber \\
    & \quad - \frac{\rho^k}{2}\|Ax^{k,r+1} + B\bar{x}^{k,r+1} + z^{k,r}\|^2 \nonumber \\
    & = -\frac{\rho^k}{2}\|z^{k,r+1}-z^{k,r}\|^2 + \rho^k\|Ax^{k,r+1} + B\bar{x}^{k,r+1} + z^{k,r+1}\|^2 \nonumber 
\end{align} 
\end{small}
\end{center}
From \eqref{eq:A1.3.2}, 
\begin{equation}\label{eq:A1.3.4}
\begin{small}\begin{aligned}
    Ax^{k,r+1} + B\bar{x}^{k,r+1} + z^{k,r+1} 
    &= \frac{1}{\rho_k}(y^{k,r+1} - y^{k,r}) \\
    &= -\frac{\beta^k}{\rho^k}(z^{k,r+1}-z^{k,r}).
\end{aligned}\end{small}
\end{equation}
Then \eqref{eq:A1.3.3} becomes
\begin{equation}\label{eq:A1.3}
    \begin{small}\begin{aligned}
    & L(x^{k,r+1}, \bar{x}^{k,r+1}, z^{k,r+1}, y^{k,r+1}) - L(x^{k,r+1}, \bar{x}^{k,r+1}, z^{k,r}, y^{k,r}) \\
    & \leq -\left(\frac{\rho^k}{2}-\frac{(\beta^k)^2}{\rho^k}\right)\|z^{k,r+1}-z^{k,r}\|^2 = -\frac{\beta^k}{2}\|z^{k,r+1}-z^{k,r}\|^2.
    \end{aligned}\end{small}
\end{equation}

\par Summing up the inequalities \eqref{eq:A1.1}, \eqref{eq:A1.2} and \eqref{eq:A1.3}, we have proved the inequality \eqref{eq:Lyapunov}. Next, we show that the augmented Lagrangian is lower bounded, and hence is convergent towards some $\underline{L}^k\in\mathbb{R}$. We note that $\upsilon(z; \lambda, \beta)$ is a convex function of modulus $\beta$, it can be easily verified that
\begin{equation}
\begin{small}\begin{aligned}
    \upsilon(z^{k,r}; \lambda^k, \beta^k) + (\lambda^k + \beta^k z^{k,r})^\top (z^\prime - z^{k,r}) \\
    + \frac{\rho^k}{2}\|z^\prime - z^{k,r}\|^2 \geq \upsilon(z^\prime; \lambda^k, \beta^k)
\end{aligned}\end{small}
\end{equation}
for any $z^\prime$, i.e.,
\begin{equation}
\begin{small}\begin{aligned}
    \upsilon(z^{k,r}; \lambda^k, \beta^k) + y^{k,r\top} (z^{k,r} - z^\prime) \\
    \geq \upsilon(z^\prime; \lambda^k, \beta^k) - \frac{\rho^k}{2}\|z^\prime - z^{k,r}\|^2.
\end{aligned}\end{small}
\end{equation}
Let $z^\prime = -(Ax^{k,r} + B\bar{x}^{k,r})$ and remove the last term on the right-hand side. Then
\begin{equation}
\begin{small}\begin{aligned}
    \upsilon(z^{k,r}; \lambda^k, \beta^k) + y^{k,r\top} (Ax^{k,r} + B\bar{x}^{k,r} + z^{k,r}) \\
    \geq \upsilon(-(Ax^{k,r} + B\bar{x}^{k,r}); \lambda^k, \beta^k).
\end{aligned}\end{small}
\end{equation}
Hence
\begin{equation}\label{eq:A1.3.5}
\begin{small}\begin{aligned}
    & L(x^{k,r}, \bar{x}^{k,r+1}, z^{k,r}, y^{k,r}) = f(x^{k,r}) + g(\bar{x}^{k,r}) + \upsilon(z^{k,r}; \lambda^k, \beta^k) \\
    &+ y^{k,r\top} (Ax^{k,r} + B\bar{x}^{k,r} + z^{k,r}) + \frac{\rho^k}{2}\|Ax^{k,r} + B\bar{x}^{k,r} + z^{k,r}\|^2 \\
    & \geq f(x^{k,r}) + g(\bar{x}^{k,r}) + \upsilon(-(Ax^{k,r} + B\bar{x}^{k,r}); \lambda^k, \beta^k). \\
\end{aligned}\end{small}
\end{equation}
Since $\upsilon(z) = \lambda^\top z+\frac{\beta}{2}\|z\|^2 \geq -\|\lambda\|^2/(2\beta)$, $\lambda$ is bounded in $[\underline{\lambda}, \overline{\lambda}]$, $\beta^k\geq \beta^1$, and $f$ and $g$ are bounded below, $L$ has a lower bound. Lemma \ref{lemma:descent} is proved.

\section{Proof of Corollary \ref{corollary:descent}}\label{app:B}
\par Taking the limit $r\rightarrow \infty$ on the both sides of inequality \eqref{eq:Lyapunov}, it becomes obvious that $B\bar{x}^{k,r+1}-B\bar{x}^{k,r}$ and $z^{k,r+1}-z^{k,r}$ converge to 0. Due to \eqref{eq:A1.3.4}, we have $Ax^{k,r} + B\bar{x}^{k,r} + z^{k,r} \rightarrow 0$. Hence there must exist a $r$ such that \eqref{eq:outer.termination} is met. At this time, the optimality conditions for $x^{k,r+1}$ is written as
\begin{equation}
\begin{small}\begin{aligned}
    0 \in & \partial f(x^{k,r+1}) + \mathcal{N}_\mathcal{X}(x^{k,r+1}) + A^\top y^{k,r} \\ &+ \rho^kA^\top (Ax^{k,r+1}+B\bar{x}^{k,r}+z^{k,r}).
\end{aligned}\end{small}
\end{equation}
According to the update rule of $y^{k,r}$, the above expression is equivalent to
\begin{equation}
\begin{small}\begin{aligned}
    0 \in & \partial f(x^{k,r+1}) + \mathcal{N}_\mathcal{X}(x^{k,r+1}) + A^\top y^{k,r+1} \\
    & -\rho^kA^\top (B\bar{x}^{k,r+1}+z^{k,r+1}-B\bar{x}^{k,r}-z^{k,r}),
\end{aligned}\end{small}
\end{equation}
i.e.,
\begin{equation}
\begin{small}\begin{aligned}
    & \rho^kA^\top (B\bar{x}^{k,r+1}+z^{k,r+1}-B\bar{x}^{k,r}-z^{k,r}) \\
    & \quad \in \partial f(x^{k,r+1}) + \mathcal{N}_\mathcal{X}(x^{k,r+1}) + A^\top y^{k,r+1}.
\end{aligned}\end{small}
\end{equation}
According to the first inequality of \eqref{eq:outer.termination}, the norm of the left hand side above is not larger than $\epsilon_1^k$, which directly implies the first condition in \eqref{eq:outer.termination.approximation}. In a similar manner, the second condition in \eqref{eq:outer.termination.approximation} can be established. The third one follows from \eqref{eq:A1.3.2} and the fourth condition is obvious.

\section{Proof of Lemma \ref{lemma:convergence.1}}\label{app:C}
\par We first consider the situation when $\beta^k$ is unbounded. From \eqref{eq:A1.3.5}, we have
\begin{equation}\label{eq:A1.4.1}
\begin{small}
\begin{aligned}
    \overline{L} \geq & f(x^{k+1}) + g(x^{k+1}) \\
    & - \lambda^{k\top}(Ax^{k+1} + B\bar{x}^{k+1}) + \frac{\beta^k}{2}\|Ax^{k+1} + B\bar{x}^{k+1}\|^2.
\end{aligned}
\end{small}
\end{equation}
Since $f$ and $g$ are both lower bounded, as $\beta^k\rightarrow\infty$, we have $Ax^{k+1} + B\bar{x}^{k+1} \rightarrow 0$. Combined with the first two conditions of \eqref{eq:outer.termination.approximation} in the limit of $\epsilon_1^k$, $\epsilon_2^k$, $\epsilon_3^k \downarrow 0$, we have reached \eqref{eq:stationary.point}. 
\par Then we suppose that $\beta^k$ is bounded, i.e., the amplification step $\beta^{k+1}=\gamma\beta^k$ is executed for only a finite number of outer iterations. According to Lines 17--21 of Algorithm \ref{alg:A1}, expect for some finite choices of $k$, $\|z^{k+1}\|\leq \omega \|z^k\|$ always hold. Therefore $z^{k+1}\rightarrow 0$. Apparently, \eqref{eq:stationary.point} follows from the limit of \eqref{eq:outer.termination.approximation}.

\section{Proof of Lemma \ref{lemma:complexity.1}}\label{app:D}
\par From Lemma \ref{lemma:descent} one knows that within $R$ inner iterations
\begin{equation}
\begin{small}\begin{aligned}
    \frac{\overline{L} - \underline{L}^k}{\beta^k} \geq \sum_{r=1}^{R} \left(\|B\bar{x}^{k,r+1} - B\bar{x}^{k,r} \|^2 + \frac{1}{2} \|\bar{z}^{k,r+1} - z^{k,r} \|^2 \right).
\end{aligned}\end{small}
\end{equation}
Then
\begin{equation}
\begin{small}\begin{aligned}
    \|B\bar{x}^{k,R+1} - B\bar{x}^{k,R} \|, \enskip \|z^{k,R+1} - z^{k,R} \| 
    \sim \mathcal{O}(1/\sqrt{\beta^k R}).
\end{aligned}\end{small}
\end{equation}
For the $k$-th outer iteration, its inner iterations are terminated when \eqref{eq:outer.termination} is met, which is translated into the following relations:
\begin{equation}
\begin{small}\begin{aligned}
    & \mathcal{O}(\rho^k/\sqrt{\beta^k R^k}) \leq \epsilon_1^k \sim \mathcal{O}(\vartheta^k), \\
    & \mathcal{O}(\rho^k/\sqrt{\beta^k R^k}) \leq \epsilon_2^k \sim \mathcal{O}(\vartheta^k), \\
    & \mathcal{O}(1/\sqrt{\beta^k R^k}) \leq \epsilon_3^k \sim \mathcal{O}(\vartheta^k/\beta^k).
\end{aligned}\end{small}
\end{equation}
where the last relation uses \eqref{eq:A1.3.4} with $\rho^k=2\beta^k$. Therefore
\begin{equation}\label{eq:A1.5.1}
\begin{small}\begin{aligned}
    R^k \sim \mathcal{O}(\beta^k/\vartheta^{2k}).
\end{aligned}\end{small}
\end{equation}
\par At the end of the $k$-th iteration, suppose that Lines 19--20 and Lines 17--18 of Algorithm \ref{alg:A1} have been executed for $k_1$ and $k_2$ times, respectively ($k_1+k_2=k$). Then the obtained $z^{k+1}$ satisfies $\|z^{k+1}\| \sim \mathcal{O}(\omega^{k_1})$, and $\|Ax^{k+1}+B\bar{x}^{k+1}+z^{k+1}\| \leq \epsilon_3^k \sim \mathcal{O}(\vartheta^k/\beta^k)$, which imply 
\begin{equation}
\begin{small}\begin{aligned}
    \|Ax^{k+1}+B\bar{x}^{k+1}\| \leq \mathcal{O}(\vartheta^k/\beta^k) + \mathcal{O}(\omega^{k_1}).
\end{aligned}\end{small}
\end{equation}
From \eqref{eq:A1.4.1}, 
\begin{equation}\label{eq:A1.5.2}
\begin{small}\begin{aligned}
    \beta^k \|Ax^{k+1}+B\bar{x}^{k+1}\|^2 \sim \beta^k(\mathcal{O}(\vartheta^k/\beta^k) + \mathcal{O}(\omega^{k_1}))^2 \sim \mathcal{O}(1).
\end{aligned}\end{small}
\end{equation}
Substituting \eqref{eq:A1.5.2} into \eqref{eq:A1.5.1}, we obtain
\begin{equation}
\begin{small}\begin{aligned}
    R^k \sim \mathcal{O}\left( \frac{1}{\vartheta^{2k}} \frac{1}{(\mathcal{O}(\vartheta^k/\beta^k) + \mathcal{O}(\omega^{k_1}))^2} \right).
\end{aligned}\end{small}
\end{equation}
When $\vartheta\leq\omega$, $\vartheta^k \leq \omega^{k} \leq \omega^{k_1}\gamma^{k_2}$, and hence $\gamma^{k_2}\vartheta^k \leq \omega^{k_1}$, i.e., $\omega^{k_1}$ dominates over $\vartheta^k/\beta^k$, leading to
\begin{equation}
\begin{small}\begin{aligned}
    R^k \sim \mathcal{O}(1/\vartheta^{2k}\omega^{2k_1}) \sim \mathcal{O}(1/\vartheta^{2k}\omega^{2k}). 
\end{aligned}\end{small}
\end{equation}
\par For $K$ outer iterations, the total number of inner iterations is
\begin{equation}
\begin{small}\begin{aligned}
    R = \sum_{k=1}^K R^k \sim \mathcal{O}\left( \sum_{k=1}^K \frac{1}{\vartheta^{2k}\omega^{2k}} \right) \sim \mathcal{O}\left(\frac{1}{\vartheta^{2K}\omega^{2K}} \right).
\end{aligned}\end{small}
\end{equation}
The number of outer iterations needed to reach an $\epsilon$-approximate stationary point is obviously $K\sim \mathcal{O}(\log_\vartheta \epsilon)$. Then 
\begin{equation}
\begin{small}\begin{aligned}
    R \sim \mathcal{O}(\epsilon^{-2(1+\varsigma)}).
\end{aligned}\end{small}
\end{equation}

\section{Proof of Lemma \ref{lemma:convergence.3}}\label{app:E}
Through the inner iterations, only Anderson acceleration might lead to an increase in the barrier augmented Lagrangian. Combining Assumption \ref{assum:3}, Assumption \ref{assum:5}, and the safeguarding criterion \eqref{eq:Lagrangian.increase.criterion}, we obtain
\begin{equation}
\begin{small}
\begin{aligned}
    & L_{b^k}(x^{k,r+1}, \bar{x}^{k,r+1}, z^{k,r+1}, y^{k,r+1}) \\
    & \quad \leq \overline{L} + \tilde{L}_0\eta_L\sum_{r=0}^{\infty} \frac{1}{r^{1+\sigma}} < +\infty,
\end{aligned}
\end{small}
\end{equation}
Together with Assumptions \ref{assum:1} and \ref{assum:2}, $L_{b^k}$ is also bounded below. Therefore $L_{b^k}$ is bounded in a closed interval and must have converging subsequences. Therefore we can choose a subsequence converging to the lower limit $\underline{L}$. For any $\varepsilon>0$ there exists an index $R$ of inner iteration in this subsequence, such that $\tilde{L}_0\eta_L \sum_{r=R}^\infty r^{-(1+\sigma)} < \varepsilon/2$ and $L_{b^k}(x^{k,r+1},\bar{x}^{k,r+1},z^{k,r+1},y^{k,r+1}) < \underline{L}+\varepsilon/2$ for any $r\geq R$ on this subsequence. It then follows that for any $r\geq R$, whether on the subsequence or not, it holds that 
\begin{equation}
\begin{small}\begin{aligned}
L_{b^k}(x^{k,r+1},\bar{x}^{k,r+1},z^{k,r+1},y^{k,r+1}) < \underline{L}+\varepsilon.    
\end{aligned}\end{small}
\end{equation}
Hence the upper limit is not larger than $\underline{L}+\varepsilon$. Due to the arbitrariness of $\varepsilon>0$, the lower limit coincides with the upper limit, and hence the sequence of barrier augmented Lagrangian is convergent. 
\par The convergence of the barrier augmented Lagrangian implies that as $r \rightarrow \infty$, $L_{b^k}(x^{k,r+1},\bar{x}^{k,r+1},z^{k,r+1},y^{k,r+1}) - L_{b^k}(x^{k,r},\bar{x}^{k,r},z^{k,r},y^{k,r}) \rightarrow 0$. 
Suppose that $r$ is not an accelerated iteration, then since this quantity does not exceed $-\beta^k\|B\bar{x}^{k,r+1} - B\bar{x}^{k,r}\|^2 - (\beta^k/2)\|z^{k,r+1}-z^{k,r}\|^2$, we must have $B\bar{x}^{k,r+1} - B\bar{x}^{k,r} \rightarrow0$ and $z^{k,r+1}-z^{k,r}\rightarrow 0$. Otherwise if inner iteration $r$ is accelerated, the convergence of $B\bar{x}^{k,r+1} - B\bar{x}^{k,r}$ and $z^{k,r+1}-z^{k,r}$ are automatically guaranteed by the second criterion \eqref{eq:norm.increase.criterion} of accepting Anderson acceleration. The convergence properties of these two sequences naturally fall into the paradigm of Lemma \ref{lemma:descent} for establishing the convergence to approximate KKT conditions of the relaxed problem.

\section*{Acknowledgment}
This work was supported by National Science Foundation (NSF-CBET). The authors would also like to thank Prof. Qi Zhang for his constructive opinions. 
\bibliographystyle{IEEEtran}
\bibliography{mybibfile}
\newpage
\begin{IEEEbiography}[{\includegraphics[width=1in,height=1.25in,clip,keepaspectratio]{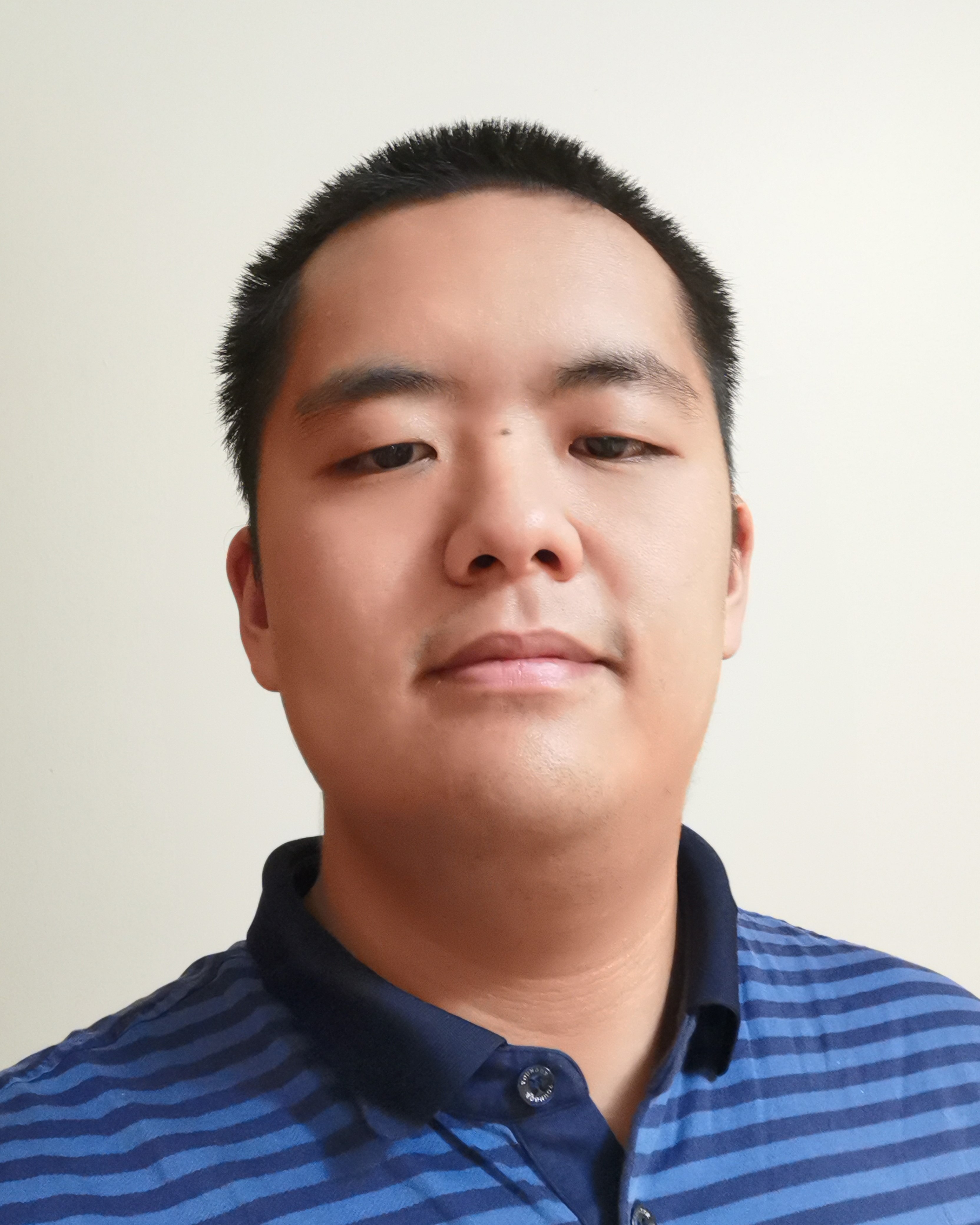}}]{Wentao Tang} was born in Yongzhou, Hunan Province, P. R. China. He received a Bachelor of Science degree in Chemical Engineering and a secondary degree in Mathematics from Tsinghua University, Beijing, China, in 2015. He is now pursuing a Ph.D. degree in Chemical Engineering at University of Minnesota. He is the recipient of the Doctoral Dissertation Fellowship of University of Minnesota for 2018--2019, and the 1st place in CAST Directors' Student Presentation Award of the 2019 AIChE Annual Meeting. His current research interests include the architecture design and algorithm of distributed and hierarchical control and optimization problems, nonlinear system identification, and data-driven control of nonlinear processes. 
\end{IEEEbiography}
\vspace{-5in}
\begin{IEEEbiography}[{\includegraphics[width=1in,height=1.25in,clip,keepaspectratio]{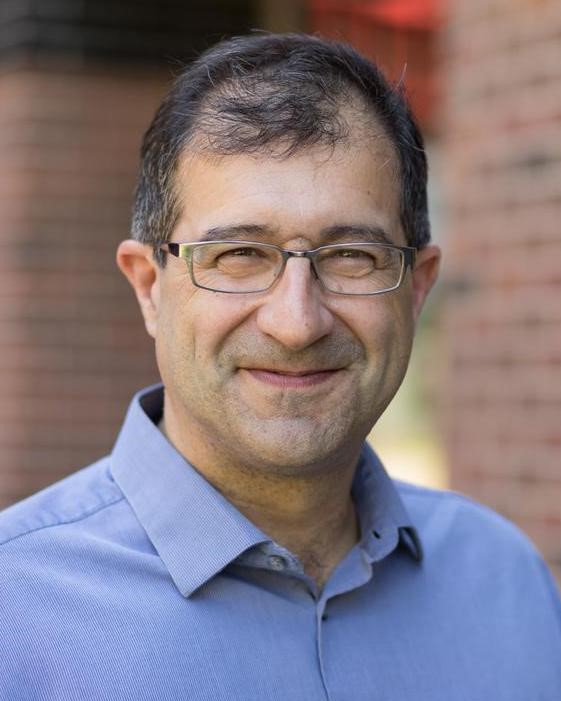}}]{Prodromos Daoutidis} is a College of Science and Engineering Distinguished Professor and Executive Officer in the Department of Chemical Engineering and Materials Science at the University of Minnesota. He received a Diploma degree in Chemical Engineering (1987) from the Aristotle University of Thessaloniki, M.S.E. degrees in Chemical Engineering (1988) and Electrical Engineering: Systems (1991) from the University of Michigan, and a Ph.D. degree in Chemical Engineering (1991) from the University of Michigan. He has been on the faculty at Minnesota since 1992, while he has also held a position as Professor at the Aristotle University of Thessaloniki (2004--06). He is the recipient of several awards and recognitions, including the AIChE Computing in Chemical Engineering Award, the PSE Model Based Innovation Prize, the Best Paper Prize from the \textit{Journal of Process Control}, an NSF Career Award, and the AIChE Ted Peterson Award. He has also been a Humphrey Institute Policy Fellow. He is the Associate Editor for Process Systems Engineering in the \textit{AIChE Journal}, and an Associate Editor in the \textit{Journal of Process Control}. He has co-authored 5 books, 290 refereed papers, and has supervised to completion 35 Ph.D. students and post-docs. His current research is on control and optimization of complex and networked systems, and the design and operation of distributed renewable systems for power generation and production of fuels and chemicals.
\end{IEEEbiography}

\end{document}